\documentstyle[twocolumn,aps,prl,epsf]{revtex}
\newcommand{\gdot}{$\dot{\gamma}$ \/}
\newcommand{\gdotl}{$\dot{\gamma} \lambda$ \/}
\newcommand{\gdotsql}{$\dot{\gamma} \sqrt{\lambda}$ \/}
\newcommand{\gdotN}{$\dot{\gamma}_N$ \/}
\newcommand{\gdotlN}{$\dot{\gamma}_N \lambda_N$ \/}
\newcommand{\gdotsqlN}{$\dot{\gamma}_N \sqrt{\lambda}_N$ \/}
\newcommand{\lN}{$\lambda_N$ \/}
\newcommand{\gNw}{$\dot{\gamma}_N/\omega$ \/}
\newcommand{\gw}{$\dot{\gamma}/\omega $ \/}
\newcommand{\si}{s$^{-1}$ \/} 
\newcommand{\beq}{\begin{equation}} 
\newcommand{\eeq}{\end{equation}} 
\newcommand{\gdotsqlNa}{$\dot{\gamma}_N \sqrt{(1 + \lambda_N^2)}$ \/}
\newcommand{\gdotsqla}{$\dot{\gamma} \sqrt{(1 + \lambda^2)}$ \/}
\newcommand{\WiR}{$\dot{\gamma} \sqrt{\lambda} \tau_R$ \/}
\newcommand{\Wid}{$\dot{\gamma} \sqrt{\lambda} \tau_d$ \/}

\begin{document} 
\twocolumn[\hsize\textwidth\columnwidth\hsize\csname
@twocolumnfalse\endcsname
\widetext
\draft

\title{Dynamics of Entangled Polymeric Fluids in Two-roll Mill studied 
via Dynamic Light Scattering and Two-color flow Birefringence. I. Steady flow}

\author{Subrata Sanyal\cite{byline}, Dmitry Yavich\cite{byline1} 
and L. Gary Leal} 
\address{Department of Chemical Engineering, University of 
California at Santa Barbara, Santa Barbara, CA 93106-5080, U.S.A.}

\date{June 13, 2000}

\maketitle


\begin{abstract}

We present the experimental results on a series of high molecular 
weight, entangled polystyrene solutions subjected to a ``mixed'' 
shear and elongational flow-type generated in a co-rotating 
two-roll mill.  For steady-flows as well as for start-up of flows 
from rest, we used two different optical methods to study the 
dynamics of entangled polymeric fluids: two-color flow birefringence 
(TCFB) and dynamic light scattering (DLS).  Using the TCFB method, we 
measured the birefringence, $\Delta n$, and the orientation angle,  
$\chi$, of the optic axes in the solution and thereby the generalized 
viscosity function, $\eta$ (with the use of stress-optical relations).  
The DLS method was applied to measure the velocity-gradient, \gdot, and 
the flow-type parameter $\lambda$ for the polymer solutions under flow 
conditions identical to the TCFB measurements.  For low deformation 
rates the symmetry of the flow-field was reduced with the use of 
polymeric fluids compared to that seen with a Newtonian fluid.  A 
molecular constitutive Doi-Edwards-Marrucci-Grizzuti (DEMG) model which 
includes polymer chain-stretching effects, has been used to numerically 
simulate the predictions for $\Delta n$, $\chi$, and $\eta$ under steady 
flow conditions for the polystyrene fluids with \gdot and $\lambda$, 
measured via DLS, as inputs to the model.  A detailed comparison of 
the DEMG model predictions with the experimental results shows that 
the model works qualitatively as well as quantitatively for the low 
and intermediate deformation rates, but fails at high rates of 
deformation by predicting a stronger chain-stretching than observed 
experimentally.  The effect of polymer molecular-weight and number 
of entanglements per chain are highlighted. \\

\end{abstract}


]

\narrowtext

\section{Introduction}

The majority of prior experimental studies of the dynamics of polymeric 
fluids were based on the simple shear flow.  The main advantage is that 
the complete kinematics of these homogeneous flow-fields is known a 
priory, and experimental measurements of velocity-gradients are not 
necessary.  On the other hand, the majority of the polymer processing 
applications employ flows that are inhomogeneous and often involve a 
mixture of simple shear and pure extension.  This have provided one 
motivation for recent experimental as well as theoretical studies of 
the behavior of the polymeric fluids in ``mixed'' or purely extensional 
flows (also called the ``strong'' flows\cite{astarita}, which in 
two-dimensions would require the magnitude of the strain-rate to exceed 
the vorticity).  Generally, such strong flows appear only as a local 
part of some otherwise globally inhomogeneous and ``weak''\cite{astarita} 
flows.  Examples of such flows include the stagnation zone in a cross-slot 
or cross-jet device; the converging zone of a contracting channel; and 
the stagnation region in two- or four-roll mills.  The velocity-field in 
these local region for any polymeric fluid will be very different from 
that of a Newtonian fluid of similar density and viscosity, because the 
flows are globally inhomogeneous and the fluid's viscoelastic nature.  
It is therefore insufficient to measure only the birefringence (or stress 
in a rheological experiment), without simultaneously measuring the 
velocity-gradient field by some technique. 

Different indirect techniques, e.g., Particle Imaging Velocimetry and 
Laser Doppler Velocimetry\cite{LDV}, hot wire anemometers\cite{balint}, 
forced Rayleigh 
scattering\cite{cloitre}, holographic grating methods\cite{garcia}, etc. 
have been employed in the past for the measurement of velocity-gradients  
in time varying flows, but each exhibits some characteristic drawbacks.  
Provided that the flow is laminar and nonchaotic, the dynamic light 
scattering (DLS) technique\cite{fuller,wang} is one extremely valuable 
method for the {\em direct} measurement of velocity-gradients in strong 
flows.  The high degree of spatial resolution of the laser light as well 
as the very high temporal resolution of the present day correlators has 
made it possible to use DLS for precise determination of velocity-gradients 
for both steady and time-dependent flows even in very small-scale flow 
systems with velocity-gradients that change relatively in time.  Further, 
unlike other methods mentioned above, DLS is nonintrusive in 
nature.  

The most frequently used indirect rheo-optical probe of polymer conformation  
in two-dimensional flows has been the flow birefringence technique\cite{J-K}.  
Since it provides an average measure across the flow in third direction, it 
is only approximately {\em local} in that sense.  Early studies using 
traditional flow birefringence technique were primarily restricted to steady 
flows.  On the other hand, recent studies\cite{dunlap,enrique,pearson1,dmitry,jim} 
for both dilute and concentrated polymer solutions in time-dependent flows 
have used two-color flow birefringence (TCFB)\cite{fuller1} and/or 
phase-modulated birefringence\cite{fuller2} techniques to simultaneously 
measure the retardance (and hence the birefringence, $\Delta n$) and the 
orientation angle, $\chi$, of the principle axis of the refractive-index 
tensor with respect to the axes fixed on the flow-cell, in a single experiment.  

The theoretical understanding of the dynamics of entangled polymer solution 
and melts is far from complete.  Using de Gennes'\cite{gennes} original idea 
of ``reptation'', Doi and Edwards (DE) proposed a tube model\cite{DE}, which 
views the polymer chain as a sequence of segments confined in a tube-like 
region.  The tube is defined by the topological constraints on the lateral 
motion and orientation imposed by neighboring chains.  The segmental stretching 
can relax on a timescale similar to that for an unconstrained chain (dilute 
systems), i.e., the Rouse time $\tau_R$, and the segmental orientations 
can only relax via reptational dynamics i.e., through longitudinal diffusion 
to escape the tube which requires a much longer timescale $\tau_d$, called 
the reptation or disengagement time.  These two time scales are related via 
$N_e$, the number of entanglement points per chain: $\tau_d = 3 N_e \tau_R$.  
The DE model is expected to be valid for highly entangled ($N_e >> 1$) 
polymeric samples.  Thus the original DE model neglects the chain-stretching 
effects by assuming that the ``snap-back'' of the stretched chain to its 
equilibrium length is instantaneous (i.e., $\tau_R << \tau_d$) and develops 
constitutive equations which incorporate segmental orientational dynamics 
only.  This simplified model has been quite successful in many rheometric 
flows such as steady simple shear and oscillatory shear, where chain-stretching 
is expected to be insignificant.  On the other hand, serious 
inherent limitations of DE theory have long been recognized for the case of 
strong, extension-dominated flows\cite{dunlap,enrique,pearson1,dmitry,jim,muller} 
and also in transient shear flows, where chain-stretching is important.  
Over the past decade, this model has seen several improvements to include 
chain-stretching in the uniform\cite{pearson1,booij}, 
non-uniform\cite{MG,pearson2} and finitely-extensible\cite{mead} form.  
The predictions of the resulting so-called Doi-Edwards-Marrucci-Grizzuti 
(DEMG) model has been studied in detail in a recent set of papers by Mead 
{\em et al.}\cite{mead1,mead2}. 

The aim of the present paper is to understand the dynamics of entangled, 
high molecular weight polymeric fluids subjected to the strong, 
extension-dominated flows (i.e., for strong two-dimensional flows in the 
sense of Astarita\cite{astarita}).  To this end, we present two-color flow 
birefringence and dynamic light scattering results for a series of entangled polystyrene 
solutions in steady, near-homogeneous, planar flows created by a co-rotating 
two-roll mill.  A detailed comparison of the experimental results for 
birefringence, orientation angle and generalized extensional viscosity 
(defined below, and obtained via stress-optical relation) for the three 
entangled samples to the predictions from the numerical simulation of 
the DEMG model\cite{mead1} is carried out using the measured flow data 
as input to the model.  The effect of molecular weight and number of 
entanglements per chain of the polymers on the results are demonstrated. 

In sec.~II we provide the necessary details related to our experiments, 
namely, the samples we used (sec.~II.A); the linear viscoelastic 
measurements (sec.~II.B); and the experimental apparatus (sec.~II.C).  
The two-color flow birefringence and the dynamic light scattering 
techniques and apparatus involved are described in brief along with the 
two-roll mill flow-cell in sec.~II.C.  Section~III deals with the results 
of the steady-state flow experiments (sec.~III.A) using the DLS method 
and comparison of the steady flow TCFB results with the DEMG model 
(sec.~III.B).  Finally, sec.~IV contains a summary of our findings and 
conclusion. 

\section{Experimental details}

\subsection{Materials}

The Newtonian fluid sample used was a suspension of spherical polystyrene 
particles (polyballs) of diameter $0.11 \mu$m (Polysciences Inc., U.S.A.) 
in glycerol (Sigma Chemical Co., U.S.A., ACS Reagent) at a concentration 
of 150 ppm.  As shown in Table\ \ref{samples}, three non-Newtonian fluids, 
named PS81, PS82 and PS2, were made using two standard, fairly monodisperse 
polystyrene samples (Tosoh Co., Japan), namely $M_w = 8.42 \times 10^6$, 
$M_w/M_n = 1.17$, Lot. No. TS-31 for one and $M_w = 2.89 \times 10^6$, 
$M_w/M_n = 1.09$, Lot. No. TS-10 for the other.  Here $M_w$ and $M_n$ 
specify the weight-average and the number-average molecular weights 
respectively.  The solvents were prepared by adding polystyrene oligomer 
with $M_w = 6000$ or $2500$ and a broad molecular weight distribution 
to toluene (Aldrich Chemical Co., Inc., U.S.A., ACS spectrophotometric 
grade) at mixing ratios by weight (toluene: 2500 PS = 43:57 for PS81 
and PS82, and 48:52 for PS2), and allowing the 
mixture to dissolve at room temperature ($= 20^\circ$C) for a few days 
with occasional slow mixing using a magnetic stirrer.  Measured amounts  
of high molecular weight polystyrene samples were then added to these 
oligomer solvents to achieve the desired concentrations, $c$, shown in 
Table\ \ref{samples}.  The complete solution was then thoroughly mixed 
in a glass bottle for several hours by continuously rolling the bottle on 
its side in a Rollacell (New Brunswick Scientific Co., Inc., U.S.A.).  
In order to reduce the background scattering in the light scattering 
experiments, the samples were filtered using Milipore filters with $5 
\mu$m pore size for the polymer solutions and $0.45 \mu$m pore size for 
the polyball suspension.

The flow-cell was pre-cleaned with acetone (Fisher Scientific, U.S.A., 
ACS spectranalyzed), oven-dried at $50^\circ$C and brought back to the 
room temperature everytime before the fluid sample was transferred into 
the cell via an air-tight siphon system.  To minimize the loss of toluene 
in the precess of fluid transfer, the use of an inactive gas in the siphon 
system was necessary.   For this purpose, we used compressed nitrogen 
gas to maintain a minimum pressure differential in the siphon to fill 
up the cell volume ($\sim 100$ mL) in a couple of hours. 

\begin{figure}
\centerline{\epsfxsize = 8cm \epsfbox{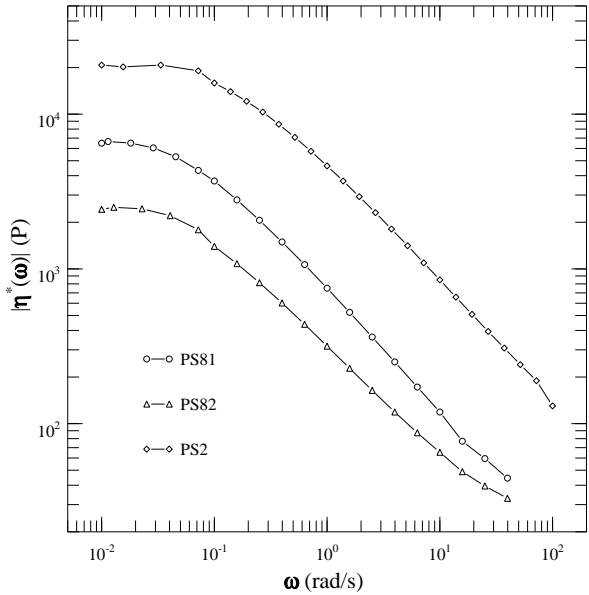}}
\caption{The modulus of complex dynamic viscosities, $|\eta^\star|$ (in P), 
as a function of the radial frequency, $\omega$ (in rad/s), for the three 
polystyrene samples, measured at $20^\circ$C.}
\label{visc} 
\end{figure}

\begin{figure}
\centerline{\epsfxsize = 8cm \epsfbox{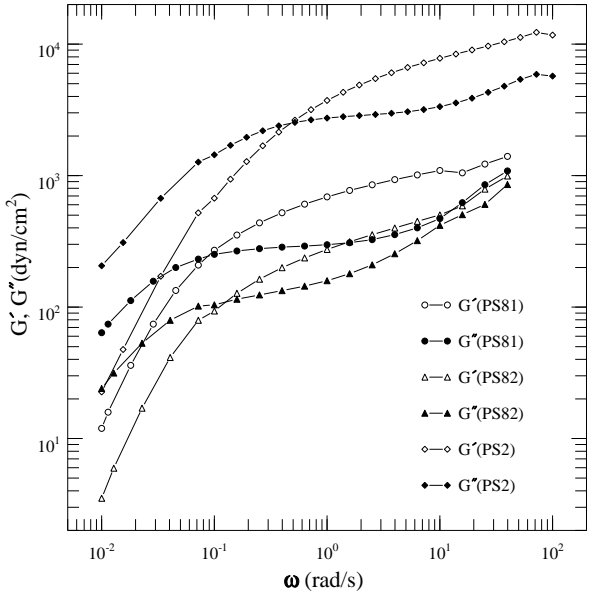}}
\caption{The dependence of the elastic modulus, $G^\prime$, and the viscous 
modulus, $G^{\prime\prime}$, (both in the units of dyn/cm$^2$) on angular 
frequency, $\omega$ (in rad/s), for the three polystyrene samples at 
$20^\circ$C.}
\label{G's} 
\end{figure}

\subsection{Linear viscoelastic measurements}

The linear viscoelastic shear flow properties of the three polymer 
solutions were measured, at a temperature of $20^\circ$C, using a 
Rheometrics Mechanical Spectrometer (RMS-800) in a cone-and-plate geometry 
with a cone diameter of 40 mm and a gap angle of $4^\circ$.  The 
measured modulii of complex dynamic viscosities $|\eta^\star|$ of the 
three liquids are plotted against the angular frequency $\omega$ in 
Fig.\ \ref{visc}.  We note that the viscosity of PS2 is about an order 
of magnitude higher than the other two samples; it is a much less mobile 
liquid.  The sample PS81 is a couple of times more viscous than PS82.  
The typical nature of the three curves seen in Fig.\ \ref{visc} is very 
similar, namely, $|\eta^\star|$ is almost constant at low frequency 
$\omega \leq 3 \times 10^{-2}$ rad \si and beyond this $\omega$ it shows 
a ``shear'' thinning behavior, following an empirical power law 
$|\eta^\star| = K \omega^n$, over the entire experimental frequency 
range, with the correlation coefficient $R^2$ being about $0.99$ or 
higher in all three cases.  The power law constant $K$ and the exponent 
$n$, obtained from a linear regression of the data are given in Table\ 
\ref{samples}.  We note that although the values of $K$ are significantly 
different in these samples, the slope of the complex dynamic viscosity 
versus angular frequency curves in a double-logarithmic plot in Fig.\ 
\ref{visc} is about same for two solutions with $N_e \sim 13$ and 
larger than the slope with $N_e \sim 7$. 

From the low frequency Newtonian plateau of these curves, the zero 
``shear-rate'' viscosity $\eta_0$ is extracted for these samples and is shown 
in Table\ \ref{samples}.  The number of entanglement points per chain, $N_e$, 
can be defined by analogy with the classical theory of rubber elasticity,  
\beq 
G_N^0 = \left(\frac{c R T}{M_e}\right) \equiv \left(\frac{c R T}{M_w}\right) 
N_e,
\label{GN0}
\eeq 
where $c$ is the polymer concentration in g cm$^{-3}$, $M_e = M_w/N_e$ is 
the molecular weight between entanglements, $R$ is the gas constant and 
$T$ is the absolute temperature.  We have used the approximate expression 
for the plateau modulus, 
\begin{equation} 
G_N^0 = 3.44 \times 10^6 \times c^{2.4} \; \mbox{dyn/cm}^2 \hspace*{1cm} 
(c \leq 0.1 \; \mbox{g cm}^{-3}),  
\label{G_N}
\end{equation} 
obtained experimentally by Osaki {\em et al.}\cite{osaki1} for polystyrene 
solutions in a good solvent.  The values obtained using this equation agrees 
reasonably well with measured values.  Thus Eqns.\ (\ref{GN0}) and (\ref{G_N}) 
leads to 
\beq 
N_e = 1.41 \times 10^{-4} \times M_w \times c^{1.4} 
\label{N_e}
\eeq 
for our experiments and the calculated values of $N_e$ for the polystyrene 
samples are shown in Table\ \ref{samples}.

The dynamic modulii of PS81, PS82 and PS2 are measured on the same rheometer 
RMS-800 with the same pair of cone and plate as for the data in Fig.\ 
\ref{visc} and are shown in Fig.\ \ref{G's}.  Both, the elastic modulus 
($G^\prime$) and the viscous modulus ($G^{\prime\prime}$) show a strong 
frequency dependence for all three cases.  The values of the modulii 
$G^\prime$ and $G^{\prime\prime}$ for PS81 is higher than that of PS82, 
both having an order of magnitude lower values than in the case of PS2.  
Each of these pair of complex modulii shows a pattern of $G^{\prime\prime}$ 
dominating over $G^{\prime}$ in the low frequency ``terminal zone'' and 
the values of $G^{\prime}$ taking over that of $G^{\prime\prime}$ in the 
higher frequency ``plateau region'', typical of polymeric fluids.  In the 
low frequency flow region, the storage modulus $G^{\prime}$ approaches a 
quadratic dependence on $\omega$ and the loss modulus $G^{\prime\prime}$ 
shows a linear dependence, again a typical behavior of polymers with a 
narrow molecular weight distribution.  At higher frequencies $G^{\prime}$ 
and $G^{\prime\prime}$ tend to converge, which may be due to the formation of 
transient entanglement network\cite{ferry}.  The value of $\omega$ at which 
the storage modulus falls below the loss modulus is highest in PS2 and lowest 
in PS81.  Assuming that this crossing point frequency corresponds to the 
inverse of the disengagement time $\tau_d$, we have estimated the values 
of $\tau_d$ to be $67.50$ s, $54.18$ s and $16.80$ s respectively for PS81, 
PS82 and PS2.  Using the prediction $\tau_R = \tau_d/(3 N_e)$ from reptation 
theory\cite{DE}, we have calculated the Rouse time, $\tau_R$, for 
each sample as reported in Table\ \ref{samples}.  To check these values, 
we can use the correlation of Menezes and Graessley\cite{M&G} 
\beq 
\tau_R = \left[\frac{6 \left(M_c\right)_0^{a-1}}{\pi^2 \rho R T}\right] 
\frac{\eta_0}{M_w^{(a - 2)} \nu^{(a - 1)b + 1}},  
\label{tauR}
\eeq 
where for polystyrene we use\cite{ferry} $(M_c)_0 = 33,000$, $a = 3.4$, 
$b = 1.3$, density $\rho = 1.07$ g cm$^{-3}$ and volume fraction $\nu = 
\rho c$.  The calculated Rouse times are $1.95$ s, $3.84$ s and $0.87$ s 
for PS81, PS82 and PS2 respectively.  It is unclear to us why the values 
calculated from Eqn.\ (\ref{tauR}) are not in closer agreement with those 
obtained here experimentally, though they are in the same ballpark.  In 
what follows, we use the experimentally determined values listed in Table\ 
\ref{samples}.  

\begin{figure}
\centerline{\epsfxsize = 8cm \epsfbox{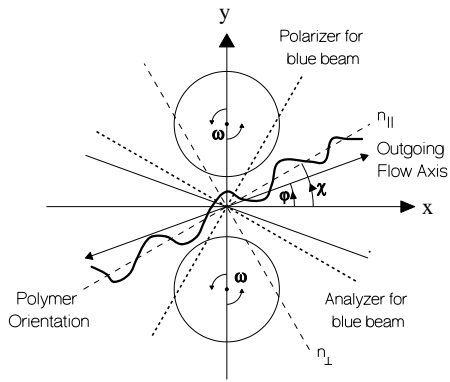}}
\caption{Relative orientation of the optical system including the 
flow-cell and the definition of the coordinate systems.}  
\label{two-roller1}
\end{figure}

\subsection{Experimental set-up} 

A schematic diagram of our experimental system is shown in Fig.\ \ref{setup}. 
It is composed of two main parts: the TCFB optical arrangement that is used 
to measure the optical anisotropy of the fluid and the DLS set-up that is 
used to measure the flow parameters of the fluid.  In the following, we 
will briefly touch upon some specific details of the apparatus relevant to 
the TCFB and DLS studies reported in this paper.  Further details of the data 
analysis technique used for TCFB\cite{enrique} and DLS\cite{wang} have been 
described elsewhere. 

\begin{figure}
\centerline{\epsfxsize = 8cm \epsfbox{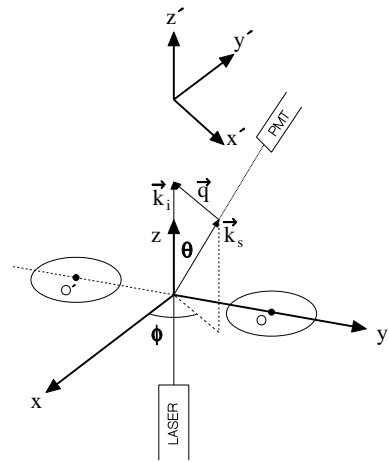}}
\caption{ Schematic representation of the light scattering configuration.}
\label{two-roller2}
\end{figure}

\subsubsection{Two-color flow birefringence}

\begin{figure}
\centerline{\epsfxsize = 8cm \epsfbox{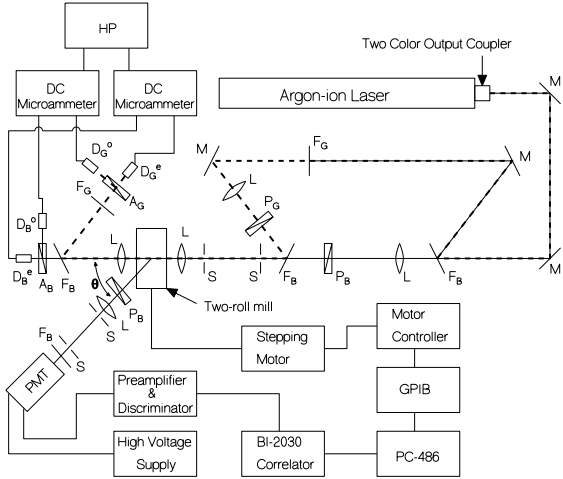}}
\caption{Schematic showing the optical layout for the two-color flow 
birefringence and the dynamic light scattering setup.  M: mirror; 
F: interference filter; L: lense; P: polarizer; A: analyzer; 
D: photodiode; S: pinhole; $\theta$: scattering angle.  The subscripts 
$B$ \& $G$ specify the blue (continuous line) and the green (broken line) 
laser beams, respectively.  The superscripts $o$ and $e$ refer to the 
ordinary and extraordinary beams, respectively.}
\label{setup} 
\end{figure}

The TCFB technique\cite{fuller1} simultaneously measures both retardance  
and orientation $\chi$ of the principal axis of the refractive-index tensor 
$\underline{\underline{\bf n}}$ relative to the axes [$(x, y)$ in Figs.\ 
\ref{two-roller1} and \ref{two-roller2}] fixed 
in the flow device.  The light source is an Argon-ion laser (Spectra 
Physics Model 2020) operating at $\sim 300$ mW, which emits two intense 
wavelengths at $\lambda_B = 4880$ \AA (blue) and $\lambda_G = 5145$ \AA 
(green).  As shown in Fig.\ \ref{setup}, with the use of several optical 
elements, each of these beams (independently polarized at $45^\circ$ 
relative to each other) are made to pass along an identical optical path 
through the sample in a two-roll mill which is placed between crossed 
polarizers.  Since the flow is non-homogeneous, to ensure the collinearity 
of the two beams (with identical optical properties, e.g., beamwidth, 
Gaussian beam-profile etc.) passing through the same element of fluid in the 
flow-cell is very crucial in this set-up.  Each of the measured intensities 
(at the photodetectors) depends upon the current degree of optical anisotropy 
of the sample in the plane of flow [$(x, y)$ plane in Fig.\ \ref{two-roller1}], 
--- i.e., the birefringence, $\Delta n = n_\parallel - n_\perp$ --- 
and $\chi$, which provides a measure of the average orientation of Kuhn 
segments in the polymeric liquid.  Further details of the optical system 
are described in Ref.\ \cite{enrique} and will thus not be repeated here.  
The measured overall error due to nonidealities of the optical components 
indicate that the maximum extinction ratios detectable in this set up with 
the flow device (loaded with polymer solution), for both colors, are 
typically ${\cal O}(2 \times 10^{-5})$. 

\subsubsection{The two-roll mill and the flow}

The two-roll mill is a flow device\cite{enrique} consisting of two 
cylinders driven simultaneously by a single DC stepping motor (Superior 
Electric, U.S.A., SLO-SYN motor type MO62-FD-09).  The cylinders rotate 
at the same angular velocity and in the same direction to generate a flow 
between the roller pair, that can be approximated at the central stagnation 
point by a linear (or homogeneous), planar velocity-field of the form 
\begin{equation} 
v = {\bf \nabla \vec{v}} \cdot \vec{r}. 
\label{v} 
\end{equation}
Here, $\vec{r}$ is the position vector defined in the flow plane w.r.t. 
the $(x, y)$ frame, shown in Figs.\ \ref{two-roller1} and \ref{two-roller2}.  
The velocity-gradient tensor ${\bf \nabla \vec{v}}$ in this coordinate system 
is given by 
\begin{equation} 
{\bf \nabla \vec{v}} = \dot{\gamma} \left[ \begin{array}{cc} 
\epsilon & 1  \\
\lambda & - \epsilon 
\end{array} \right], \hspace*{1cm} 0 < \lambda \leq 1. 
\label{nablav} 
\eeq
If we assume further that the flow is {\it symmetric} about the central 
$(x, z)$ plane between the rollers, as in the case for a Newtonian fluid at 
zero Reynolds number, then 
\begin{equation} 
\epsilon \equiv 0 \hspace*{1cm} (\hbox{symmetric flow}).  
\label{symm}
\eeq
In that case, Eqn.\ (\ref{nablav}) can be completely characterized by two 
scaler parameters, namely, the magnitude of the velocity-gradient \gdot 
$\equiv |{\bf \nabla \vec{v}}|$ (also called the ``shear-rate'') and the 
flow-type parameter $\lambda$.  The parameter $\lambda$ is a measure of 
the ratio of strain-rate to vorticity, and is defined as, 
\beq 
\frac{\|\bf{\underline{\underline{E}}}\|}{\|\bf{\underline{\underline{\Omega}}}\|} 
= \frac{1 + \lambda}{1 - \lambda}.  
\label{ratio} 
\eeq
Thus, $\lambda = 0$ corresponds to a simple shear flow, $\lambda = 1$ to 
a purely extensional (also called hyperbolic) flow, and intermediate values 
of $\lambda$ represent a ``mixed'' shear and elongational flow-type (also 
called a ``strong'' or extension dominated flow\cite{astarita}, since the 
strain-rate exceeds the vorticity).  In sharp contrast to other extensional 
flow devices, polymer molecules at the stagnation region of a two-roll mill 
have very long residence time and hence are subjected to very large total 
strains.  Consequently, they can, in principle, achieve the maximum change 
in conformation that is consistent with a particular strain-rate\cite{dunlap}.  
For a Newtonian fluid, at very low Reynolds numbers, where the symmetric 
flow assumption is valid, the creeping flow solutions\cite{dunlap} for a 
two-roll mill in an unbounded fluid approximately relate the values of the 
corresponding velocity-gradient \gdotN and the flow-type parameter 
$\lambda_N$ in the stagnation region with the geometry of the cell (i.e., 
the roller radius $R$ and the gapwidth $2h$), via   
\begin{eqnarray} 
\lambda_N = \left(\frac{4 \coth K_f}{K_f} - 1\right)^{-1}, \; 
\dot{\gamma}_N = \frac{A\omega}{\lambda_N}    \nonumber  \\
 \hbox{with} \; A = \frac{K_f}{K_f + \sinh K_f \cosh K_f}, 
\label{creep} 
\end{eqnarray}
where $\omega$ is the angular velocity of the rollers and the parameter $K_f$ 
is given by 
\beq 
1 + \frac{h}{R} = \cosh K_f.  
\label{creep1}
\eeq
The symmetry axes of the flow-field, the principal optical axes of the solution, 
as well as the relative orientation of the blue polarizer and analyzer for the 
two-roll mill set is shown in Fig.\ \ref{two-roller1}.  It may be noted that 
the Newtonian value of the flow-type parameter, $\lambda_N$, is also related to 
the acute angle of crossing, $2\varphi$, of the streamlines passing through the 
stagnation point, according to 
\beq 
\lambda_N = \tan^2 \varphi.
\label{creep2}
\eeq
The present two-roller can, in general, be used\cite{enrique} with a pair of 
rollers chosen from a set of eight such pairs of different diameters covering 
$0 \leq \lambda \leq 0.25$.  For this study, the dimensions of the rollers used 
are specified in Table\ \ref{two-roll}, which corresponds to $\lambda = \lambda_N 
= 0.1501$ for a Newtonian fluid.  The subscripts ``th'' and ``exp'' in Table\ 
\ref{two-roll} refer to the theoretical and experimental values, respectively.  
Eqns.\ (\ref{creep}) and (\ref{creep1}) allow us to estimate the theoretical 
value of $(\dot{\gamma}_N/\omega)_{\mbox{th}}$ for this set of rollers, as given 
in Table\ \ref{two-roll}.  

The stepping motor is interfaced to a computer (Hewlett Packard, model 9133 
for TCFB or PC-486 for DLS) via a GPIB (National Instruments, U.S.A., model 
NI-488.2) switch board as shown in Fig.\ \ref{setup}.  The motor has a 
fast response time of ${\cal O}(10$ ms) and a maximum acceleration of 
100 000 steps s$^{-2}$.  Gears with reduction ratios $5:1$ or $20:1$ are 
used between the motor and the flow device to cover a wide range of 
velocity-gradients.  The accuracy involved in repositioning of the flow 
cell with respect to the incident beam, even when it is dismounted for 
the purpose of replacing the solution, is always better than $0.0025$ cm. 
in $x$, $y$ and $z$ directions (as defined by three translation stages 
used to control the respective movements) and $0.001$ rad in the azimuthal 
orientation $\phi$ of the flow-cell (defined by the micrometer used to 
control the orientation) [Fig.\ \ref{two-roller2}].  The flow-cell was 
thermostated within $\pm 0.2^\circ$C via a temperature-regulated waterbath 
circulator.

\subsubsection{Dynamic light scattering}

The optical setup for the dynamic light scattering experiment was built 
around the TCFB apparatus, as can be seen from Fig.\ \ref{setup}, by mounting 
the necessary optical accessories on a triangular optical rail placed 
on the rotating arm of a massive goniometer that defines the scattering angle, 
$\theta$, and hence the scattering vector $\vec{q}$ [$q \equiv |\vec{q}| = 
\frac{4 \pi n}{\lambda_B} \sin(\theta/2)$ as shown in Fig.\ \ref{two-roller2}, 
where $n$ is the refractive-index of the solution].  The green beam of the 
laser is blocked using a beam stop near P$_G$, in Fig.\ \ref{setup}.  The 
incident blue beam scattered by the sample is polarized and collimated to 
project an image of the scattering volume at the photomultiplier tube (PMT) 
[Hamamatsu, model R647-04].  A blue line filter, F$_B$, obstructs any spurious 
light from reaching the detector.  The preamplified and discriminated PMT 
pulses are fed to a 72-channel correlator (Brookhaven Instruments, U.S.A., 
model BI-2030) to construct time-resolved intensity autocorrelation functions.  
The commercial correlator software has been modified and used in PC-486 
computer to control the flow experiment, as well as to analyze the correlation 
functions\cite{wang}.  

Provided that the seed particles are isotropic scatterers, and the time scale 
associated with the velocity-gradient [$t_{\dot{\gamma}} \sim (q \dot{\gamma} 
L)^{-1}$, where $L$ is the laser beamwidth] is much smaller than the time 
scale for diffusive motion [$t_D \sim (Dq^2)^{-1}$], the homodyne intensity 
correlation function for a general linear flow in the beam coordinates 
$(x^\prime, y^\prime, z^\prime)$ [Fig.\ \ref{two-roller2}] is given 
by\cite{fuller1,wang} 
\beq
F_2(\vec{q}, t) = \beta^\prime \left| \int \int \int d^3\vec{r}^\prime 
I(\vec{r}^\prime) \exp \{-i \vec{q}^\prime \cdot {\bf \nabla \vec{v}^\prime} 
\cdot \vec{r}^\prime t \} \right|^2.   
\label{F2} 
\eeq
Here, $\beta^\prime$ is the coherence factor defined by the optical 
geometry\cite{berne}.  The velocity-gradient tensor ${\bf \nabla \vec{v}}$ 
of Eqn.\ (\ref{nablav}) can be expressed in the beam coordinates as 
\begin{equation} 
{\bf \nabla \vec{v}^\prime} = {\bf Q \cdot \nabla \vec{v} \cdot Q}^T 
\mbox{ where } {\bf Q} = \left[ \begin{array}{cc} 
\cos \phi  &  -\sin \phi   \\
\sin \phi  & \cos \phi  
\end{array} \right].   
\label{Q} 
\end{equation}
In Fig.\ \ref{two-roller2}, the beam coordinates are chosen such that the 
scattering vector $\vec{q}^\prime$ is orthogonal to the $y^\prime$ axis, i.e., 
\beq
\vec{q}^\prime \cdot {\bf \nabla \vec{v}^\prime} \cdot \vec{r}^\prime = 
q_x f(\phi) x^\prime + q_x g(\phi) y^\prime, 
\label{q'} 
\eeq
where $q_x = q \cos(\theta/2)$, $f(\phi) = \cos \phi (\epsilon \cos \phi 
- \dot{\gamma} \lambda \sin \phi) - \sin \phi (\dot{\gamma} \cos \phi + 
\epsilon \sin \phi)$ and $g(\phi) = \sin \phi (\epsilon \cos \phi - 
\dot{\gamma} \lambda \sin \phi) + \cos \phi (\dot{\gamma} \cos \phi + 
\epsilon \sin \phi)$. 

Using a Gaussian intensity profile, Eqns.\ (\ref{Q}) and (\ref{q'}) 
in Eqn.\ (\ref{F2}), we get 
\beq
F_2 (\vec{q}, t) = \beta \exp\left\{-\frac{1}{2} L^2 q^2 t^2
\cos^2\left(\frac{\theta}{2}\right) h(\phi)\right\},  
\label{F2t}
\eeq
where $\beta \propto \beta^\prime L^6 \csc^2 \theta$ and  
\beq 
h(\phi) = f^2(\phi) + g^2(\phi).
\label{hphi}
\eeq
The values of $h(\phi)$ are evaluated in Table\ \ref{tab1} for three 
different azimuthal orientations of the two-roll mill.  Thus, with the 
assumption of a symmetric flow [Eqn.\ \ref{symm}], \gdot can be obtained from 
the decay rate $h(\phi)$ of the measured correlation function [Eqn.\ 
(\ref{F2t})] at one orientation, $\phi = 0$, of the flow-cell and 
$\lambda$ from an additional measurement at a second orientation, 
$\phi = 90^\circ$ (see, Table\ \ref{tab1}).  The correlation 
functions were accumulated only after the flow had attained its 
steady value for the corresponding motor speed.  To improve the signal 
to noise ratio, each correlation function was obtained by averaging 
over many repetitions of the experiment, ranging from a minimum of 
200 repetitions at the high roller speeds, up to a maximum of 800 
repetitions at the low roller speeds.  Using a simulated annealing 
Monte Carlo fitting procedure\cite{ss}, these average correlation 
functions were then fitted to Eqn.\ (\ref{F2t}).  In order to verify 
that the experimentally obtained correlation functions are very close 
to exponential in nature so that the above procedure followed to 
extract the velocity-gradient components from the decay time is indeed 
justified, we required a correlation coefficient $R^2$, specifying the 
quality of the fit, exceeding $0.99$.  

Contrary to the basic assumption of optical isotropy of scatterers 
used to derive Eqn.\ (\ref{F2t}), Wang et al\cite{wang} had shown 
that this equation could also be used for the scattering from 
(intrinsically anisotropic) polymer molecules.  In that case, the 
pre-exponential factor $\beta$ is shown to be directly related to 
the components of the intrinsic molecular polarizability tensor.  

\begin{figure}
\centerline{\epsfxsize = 8cm \epsfbox{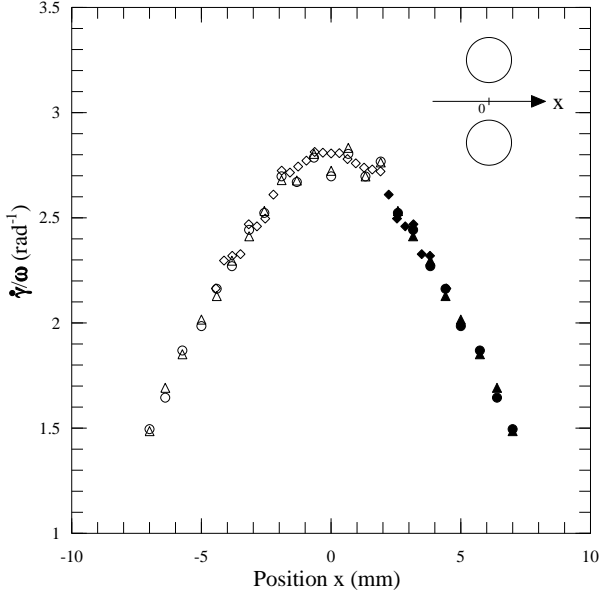}}
\caption{The velocity-gradient profile along the central crossing line 
of the rollers with $\lambda = 0.1501$ and $\phi = 0^\circ$ for the 
Newtonian fluid at $\omega = 1.07$ rad/s (triangles and circles) and 
$\omega = 1.57$ rad/s (diamonds).  The stagnation point was defined 
as the zero position.  The hollow symbols denote the measured data and 
the filled symbols denote the data obtained by symmetry.}
\label{gprofile}
\end{figure}

\section{Results}

\subsection{DLS steady flow results} 

\subsubsection{Flow characterization}

The DLS technique provides us the opportunity to map the flow-field in 
a large region between the rollers.  In order to look for changes in 
the steady flow-field due to viscoelasticity in the presence of polymers 
compared to that seen with a Newtonian fluid, we first characterize the 
flow using a Newtonian fluid.  This provides a comparison with predictions 
from the creeping flow theory for an unbounded two-roll mill.  The 
measured velocity-gradient \gdot along the $x$ axis through the stagnation 
point of the two-roll mill orientated at $\phi = 0^\circ$ for an angular 
velocity $\omega = 1.57$ rad/s is plotted along with two sets of data 
obtained by Wang et al\cite{wang}, in terms of \gw in Fig.\ \ref{gprofile}.  
The filled symbols denote the points where the geometric construction of 
the flow-cell did not allow measurements, and those points are therefore 
obtained by the symmetric reflection of the hollow measured points at 
negative $x$ values.  As can be clearly seen from this figure, different 
sets of experimental results obtained at widely separated times with 
different roller rotation rates overlap perfectly.  In a small region 
surrounding the stagnation point the velocity-gradient were approximately 
constant, \gw = $2.8$, certifying that the flow-field in this region is 
nearly homogeneous.  Also, this value compares very well with the 
theoretically expected value of \gNw = $2.7$ (Table.\ \ref{two-roll}).  
From here onwards, we shall use \gdotN and \lN in the text to specify the 
values of $\dot{\gamma}$ and $\lambda$ [Eqns.\ (\ref{creep}) and 
(\ref{creep1})] at the stagnation point 
of a two-roll mill filled with a Newtonian fluid, and the imposed motor 
speed will be measured in terms of the corresponding Newtonian strain-rate 
\gdotsqlN.  

\begin{figure}
\centerline{\epsfxsize = 8cm \epsfbox{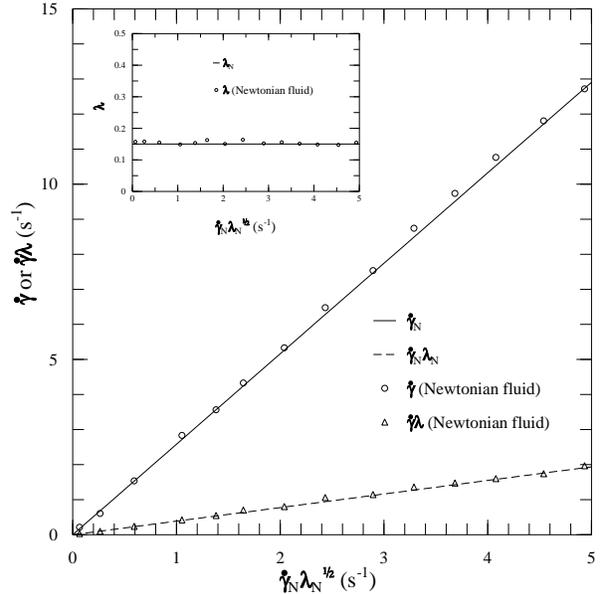}}
\caption{Linear dependence of the magnitude of the velocity-gradient 
components, $\dot{\gamma}$ and $\dot{\gamma} \lambda$ (both in the units 
of s$^{-1}$ and measured at $\phi = 0^\circ$ and $\phi = 90^\circ$, 
respectively), on the Newtonian rates of deformation, $\dot{\gamma}_N 
\lambda^{1/2}_N$ (in s), for a particular set of the rollers, $\lambda_N 
= 0.1501$, in steady flow at the stagnation region for the Newtonian fluid.  
The inset shows the strain-rate dependence of the flow-type parameter, 
$\lambda$, obtained by dividing one set of data by the other.  The 
straight lines represent the creeping flow solutions.}
\label{steadyN} 
\end{figure}

In Fig.\ \ref{steadyN}, we show the theoretical predictions for 
\gdotN and \gdotlN [Eqn.\ (\ref{creep}), (\ref{creep1}) and Table\ 
\ref{two-roll}] versus \gdotsqlN in the form of straight lines.  The 
experiments were carried out for 14 different Newtonian strain-rates, and 
for both parallel and perpendicular orientations of the sample cell.  To 
extract the experimental values of \gdot and \gdotl from the measured decay 
rates of the correlation function of Eqn.\ (\ref{F2t}) at $\phi = 0^\circ$ 
and $90^\circ$ respectively, we required a value for the beamwidth $L$.  With 
an initial guess of $L = 29$ $\mu$m, that was reported for a previous set 
of experiments\cite{wang} from this laboratory, we used a simulated 
annealing Monte Carlo fitting technique\cite{ss} to find via the fit 
shown in Fig.\ \ref{steadyN} that $L = 32$ $\mu$m yields the best match 
of the theoretical straight lines for both \gdot and \gdotl over the 
entire range of experimental strain-rates.  Compared to the previous 
experiments\cite{wang}, we have performed the present experiments 
including much higher motor speeds.  It is clearly seen in the figure 
that the corresponding data agrees well with the creeping flow solution.  
The inset shows that the extracted value of the flow-type parameter 
$\lambda$ also maintains a constant value as expected. 

\begin{figure}
\centerline{\epsfxsize = 8cm \epsfbox{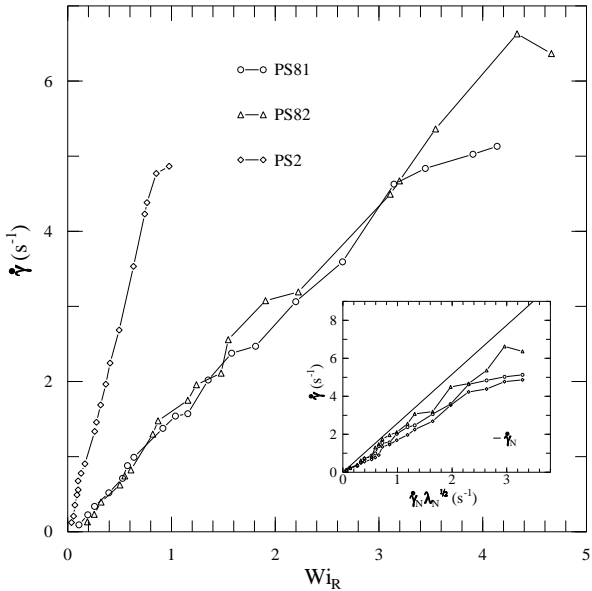}}
\caption{The dependence of the velocity-gradient, $\dot{\gamma}$ (in s$^{-1}$), 
on the Weissenberg number $Wi_R = \dot{\gamma} \lambda^{1/2} \tau_R$ (based on 
the Rouse time, $\tau_R$ and the measured $\dot{\gamma}$ \& $\lambda$), for 
$\lambda_N = 0.1501$, in steady flow at the stagnation region for the three 
entangled polystyrene fluids.  In the inset, the same data are plotted versus 
the Newtonian rates of deformation, $\dot{\gamma}_N \lambda^{1/2}_N$ (in 
s$^{-1}$), and are compared with the creeping flow solution for 
$\dot{\gamma}_N$ (straight line).}
\label{pssteady1} 
\end{figure}

In a similar manner, the flow parameters are measured for the three 
polymeric samples, PS81, PS82, and PS2, subjected to steady-state flow 
conditions.  The results are plotted in Figs.\ \ref{pssteady1}, 
\ref{pssteady2}, and \ref{pssteady3} versus the non-dimensionalized 
deformation rate, $Wi_R =$ \WiR, called the Weissenberg number (based on 
the Rouse time, $\tau_R$, and {\em measured} values of \gdot and $\lambda$).  
The Weissenberg number is defined as the ratio of a characteristic relaxation 
time for the polymeric fluid to a characteristic time for deformation.  It 
specifies the ability of a flow to generate departures of the polymer 
configuration from its static equilibrium value.  The same data are also 
plotted against \gdotsqlN, in the inset of these figures, for a comparison 
with the Newtonian values shown by the straight lines.  As expected, there 
is a substantial reduction in both \gdot and \gdotl from the corresponding 
Newtonian value.  The deviation from the Newtonian values increases with the 
increasing roller rotation rate.  Somewhat unexpectedly, however, both 
\gdot and \gdotl still appear to increase approximately linearly with 
the Newtonian strain-rate.  It is also noteworthy that this increase 
is very similar for the two samples, PS81 and PS2, which have the 
similar number of entanglements ($N_e \sim 13$) per chain (see, Table\ 
\ref{samples}).  On the other hand, PS82, with $N_e \sim 7$ shows a 
different slope for its almost linear increment with increasing 
\gdotsqlN.  When plotted against the Weissenberg number (Figs.\ 
\ref{pssteady1} and \ref{pssteady2}), the qualitative feature of the curves 
for \gdot and \gdotl appear to be quite different than that against 
\gdotsqlN: they become much more linear and owing to an order of magnitude 
smaller $\tau_R$, the increment for PS2 with the increasing rate of 
deformation is much steeper than the other two samples.  Also, \gdot versus 
$Wi_R$ data for the two samples with similar molecular weight but different 
$N_e$ are similar up to $Wi_R \sim 3$.  The quantitative similarity in 
deformation rate dependence for the samples with similar $N_e$ is more 
apparent for the flow-type parameter $\lambda$, as can be seen in Fig.\ 
\ref{pssteady3}.  The qualitative nature of this dependence for all 
three samples appears to be similar, namely, it reaches an almost constant 
value ($\sim 0.1$ for the samples with $N_e \sim 13$ and $\sim 0.05$ for 
PS82) at high deformation rates $Wi_R \geq 0.4$ (\gdotsqlN $\geq 1$) 
and at low rates of deformation exceeds the Newtonian values.  There 
is an intermediate transition regime between these these two different 
behaviors.  The parameter $\lambda$ attaining values higher than the 
Newtonian value at the lowest $Wi_R$ studied, seems surprising since 
in the limit of extremely low deformation rate ($Wi_R \rightarrow 0$), 
one should expect even a Non-Newtonian fluid to exhibit Newtonian 
behavior.  We are still away from this limit in that the Weissenberg 
number corresponding to the lowest motor speeds for the steady flow 
experiments are $\sim 0.05$ (PS2), $\sim 0.1$ (PS81) and $\sim 0.2$ 
(PS82).  As mentioned earlier, each data point in \gdot and \gdotl 
curves are obtained by averaging over several repeated experiments 
and then the $\lambda$ values are obtained by point to point division 
of the two.  Following a similar procedure, we have calculated the 
error-bars on the $\lambda$ values from the standard deviations of 
the repeated experiments for \gdot and \gdotl and plotted in Fig.\ 
\ref{errorbar}.  As expected, the error-bars increase for lower $Wi_R$, 
because of the division of smaller values of the corresponding \gdotl 
by \gdot.  Even within the limits of these error-bars we can clearly 
see that $\lambda$ initially exceeds \lN.  We have confirmed that 
this finding is not an experimental artifact by two other experimental 
means employed in this study, namely, the measurements of birefringence 
and flow parameters in transient flow conditions, and is described 
in section~III.A.3 of this paper.  

\begin{figure}
\centerline{\epsfxsize = 8cm \epsfbox{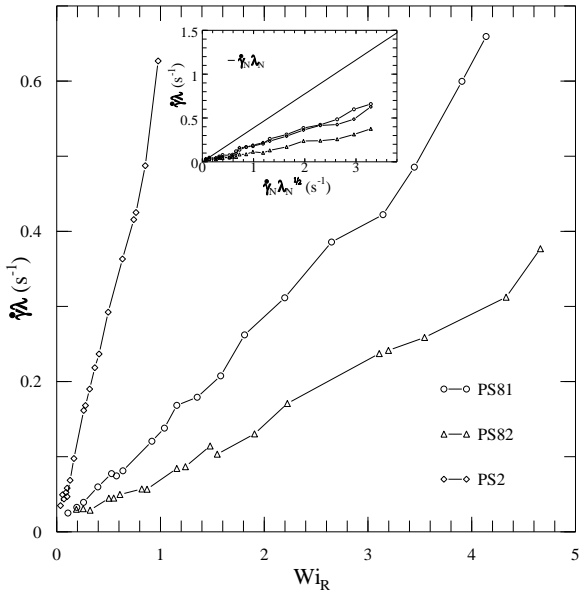}}
\caption{The dependence of the component of velocity-gradient measured at 
$\phi = 90^\circ$, i.e., $\dot{\gamma} \lambda$ (in s$^{-1}$), on the 
Weissenberg number $Wi_R = \dot{\gamma} \lambda^{1/2} \tau_R$, for 
$\lambda_N = 0.1501$, in steady flow at the stagnation region for the three 
entangled polystyrene fluids.  In the inset, the same data are plotted versus 
the Newtonian rates of deformation, $\dot{\gamma}_N \lambda^{1/2}_N$ (in 
s$^{-1}$), and are compared with the creeping flow solution for 
$\dot{\gamma}_N \lambda_N$ shown by the straight line).}
\label{pssteady2} 
\end{figure}

\begin{figure}
\centerline{\epsfxsize = 8cm \epsfbox{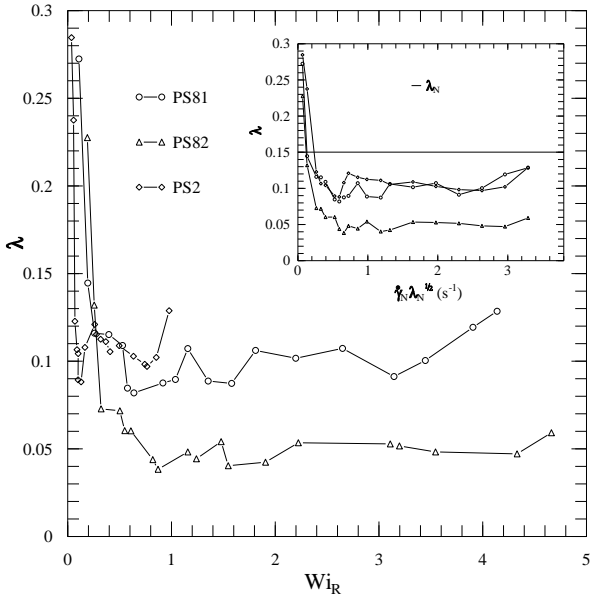}}
\caption{The flow-type parameter $\lambda$ versus $Wi_R$ for $\lambda_N = 
0.1501$, in steady flow for the three entangled polystyrene fluids, extracted 
by dividing the data in Fig.~8 by the corresponding data in Fig.~7.  The inset 
shows the same data, plotted against $\dot{\gamma}_N \lambda^{1/2}_N$ (in 
s$^{-1}$) and compared with the creeping flow solution of $\lambda_N = 
0.1501$ (straight line).} 
\label{pssteady3} 
\end{figure}

\begin{figure}
\centerline{\epsfxsize = 8cm \epsfbox{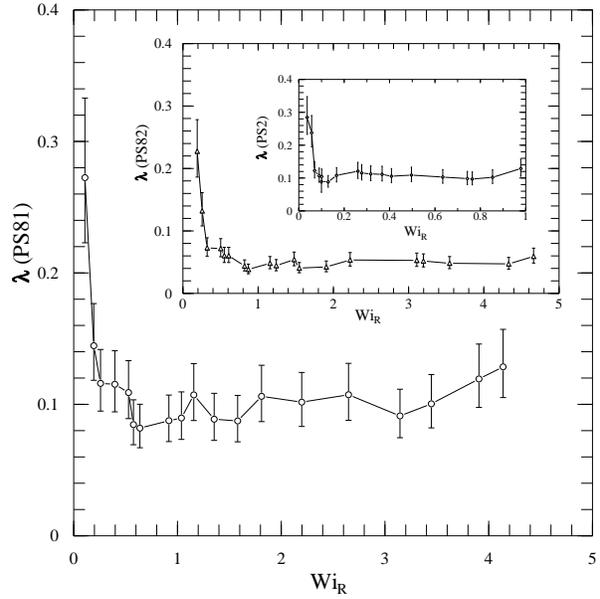}}
\caption{The error-bars represent the standard deviation on the flow-type 
parameter, $\lambda$, measured at several Weissenberg numbers, $Wi_R$, for 
the steady-state flow of the entangled polymeric solutions.}   
\label{errorbar} 
\end{figure}

\begin{figure}
\centerline{\epsfxsize = 8cm \epsfbox{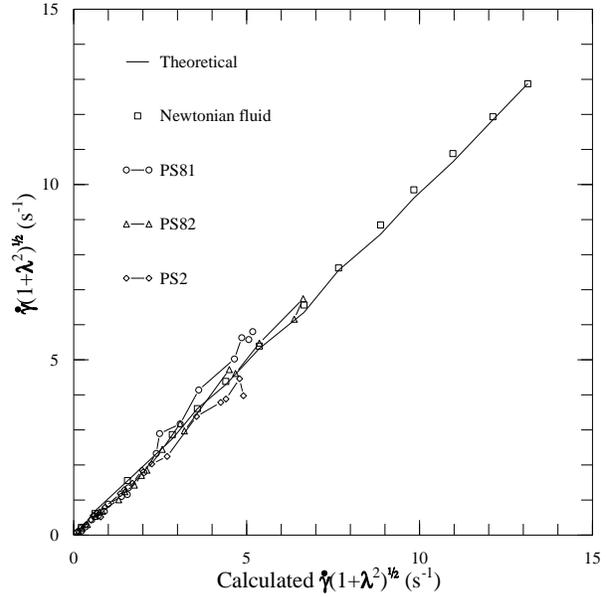}}
\caption{The connected symbols illustrate the dependence of the velocity 
gradient component measured at $\phi = 45^\circ$, i.e., $\dot{\gamma} 
(1 + \lambda^2)^{1/2}$, against its expected value, calculated using 
$\dot{\gamma}$ and $\lambda$ values measured at $\phi = 0^\circ$ and 
$\phi = 90^\circ$, respectively, for the three entangled polystyrene 
fluids in steady flow at the stagnation region.  In a similar fashion, 
the experimental and the theoretical (creeping flow) data for the 
Newtonian fluid are shown by the hollow squares and the solid line, 
respectively.} 
\label{symmetry} 
\end{figure}
\subsubsection{Flow symmetry}

As we have mentioned before, the assumption that the flow in the 
two-roll mill is symmetric about the plane passing exactly midway 
between the co-rotating rollers $(x, z)$ plane in Figs.\ 
\ref{two-roller1} and \ref{two-roller2}) 
is valid for the creeping flow of a Newtonian fluid. 
For a viscoelastic fluid, one might expect the symmetry of the flow 
to change compared to the Newtonian case.  Using the measured values 
of \gdot and \gdotl (Figs.\ \ref{pssteady1} \& \ref{pssteady2}) at 
several different steady Newtonian strain-rates, we have calculated 
the expected values of \gdotsqla.  Then, by repeating exactly same 
steady flow conditions that were used to measure \gdot and $\lambda$, 
we have extracted the velocity-gradient component \gdotsqla from the 
decay rates of the correlations functions measured directly with the 
two-roll mill oriented at $\phi = 45^\circ$ (see, Table\ \ref{tab1}).  
The result is plotted versus the corresponding calculated values at 
the same Newtonian strain-rates with the connected hollow symbols in 
Fig.\ \ref{symmetry} for the three viscoelastic samples.  The hollow 
squares represent the measured values of \gdotsqla for the Newtonian 
fluid at $\phi = 45^\circ$.  Both these and the theoretical values, 
i.e., \gdotsqlNa (shown by the solid line) at several strain-rates 
are plotted in Fig.\ \ref{symmetry} against the expected values of 
the same parameter that are calculated using the measured \gdot and 
\gdotl data from Fig.\ \ref{steadyN}.  If the symmetry of the flow 
is maintained, then Eqn.\ (\ref{F2t}) should be strictly valid for 
all orientations, $\phi$, of the flow-cell.  This means that each 
curve in Fig.\ \ref{symmetry} should be linear.  As can be clearly 
seen from the figure, this is best followed for the Newtonian fluid 
indicating the fact that the flow is symmetric in this case, as 
expected.  On the other hand, the data for the entangled fluids 
indicate that the flow symmetry is maintained only approximately, i.e., 
within the experimental accuracy of $10 \%$, for the first five or 
six data points (i.e., for \gdotsqlN $< 0.5$ \si).  The data for PS81 
and PS2 (both with $N_e \sim 13$) deviate more from the linearity 
with the increase of \gdotsqlN, but for PS82 ($N_e \sim 7$) the 
experimental data are fairly close to a straight line over the 
entire range of strain-rates studied.  At high rates of 
deformation, the measured \gdotsqla is higher than the expected 
value for PS81 and PS82, but is lower in the case of PS2. 

In order to quantify how flow symmetry changes with the deformation 
rate, we now proceed to evaluate the parameter $\epsilon$ of Eqn.\ 
(\ref{nablav}).  To do this, we can first extract the values for 
$h(\phi = 0^\circ)$, $h(\phi = 45^\circ)$ and $h(\phi = 90^\circ)$ 
from the experimentally measured decay rates of the correlation 
function at the three aforementioned orientations of the two-roll 
mill.  When these values are used in conjunction with Table\ 
\ref{tab1}, we get three equations involving the three unknowns 
\gdot, $\lambda$ and $\epsilon$.  We have used the Newton-Raphson's 
method\cite{numrec} to numerically solve these equations for the 
case of a Newtonian fluid as well as for a representative case of 
the viscoelastic fluids of our study, namely, for PS2.  The results 
are shown in Fig.\ \ref{eps}.  Our experiments with a Newtonian 
fluid shows that $\epsilon$ is zero, i.e., the flow is perfectly 
symmetric, for strain-rates upto about 2 \si.  Then, as the 
deformation rate is increased the value of $\epsilon$ increases 
almost linearly, though its value remains extremely small ($\sim 
10^{-5}$) even at the highest strain-rates studied, justifying the 
fact that, for a Newtonian fluid the flow-symmetry is maintained. 
This also explains why the first seven data points (\gdotsqlN 
$< 2$ \si) for the Newtonian fluid in Fig.\ \ref{symmetry} lies 
on the theoretical solid line and then slightly deviates from the 
line for the higher values of \gdotsqlN.  For a non-Newtonian fluid, 
on the other hand, the flow-symmetry is maintained for only up to 
\gdotsqlN $< 0.5$ \si.  Our transient flow experiments with these 
polymer solutions are carried out in this range of deformation 
rates (which corresponds to $Wi_R < 1$) and will be reported in a 
subsequent paper\cite{pst2}. 

\begin{figure}
\centerline{\epsfxsize = 8cm \epsfbox{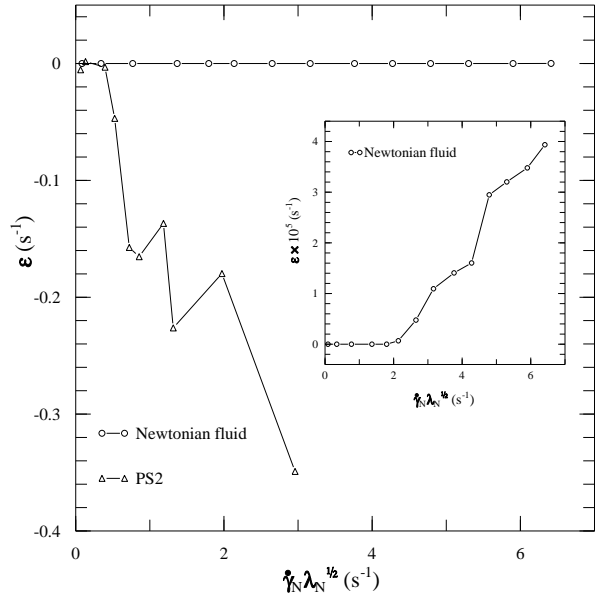}}
\caption{The parameter $\epsilon$ versus the Newtonian strain-rate, 
$\dot{\gamma}_N \lambda^{1/2}_N$, for a representative polystyrene fluid 
PS2 and the Newtonian fluid.  The inset shows the ``zoomed up'' result for 
the Newtonian fluid.}
\label{eps} 
\end{figure}

\subsubsection{Steady flow and flow-type parameter}

Let us now look into the effect of the polymer solutions on the 
steady-state values of the flow-type parameter $\lambda$ at different angular 
velocities of the rollers.  We have briefly referred to this point in 
Fig.\ \ref{pssteady3}.  Since, we have also performed extensive DLS and 
TCFB experiments on these samples subjected to transient flows\cite{pst2}, 
namely, the startup flows from rest, for several constant Weissenberg 
numbers ($Wi_R < 1$), it would be worthwhile to check how the asymptotic 
values of $\lambda$ extracted from each of those different experimental 
techniques vary with $Wi_R$ and also how they compare with the steady 
flow results of Fig.\ \ref{pssteady3}.  From the transient DLS experiments, 
we can directly get the asymptotic values of $\lambda$ but for the TCFB 
experiments, we should follow an indirect method to calculate $\lambda$.  
This can be done because for $Wi_R < 1$ the flow retains its symmetry, 
as we have shown before.  The steady-state orientation of the principal 
optic axis at appreciable strain-rates is expected to approach the outflow 
axis of the flow, or, in other words, $\chi \rightarrow \chi^\prime \approx 
\varphi$ [see, Fig.\ \ref{two-roller1}].  Hence, $\lambda$ may be 
calculated from the asymptotic orientation angle, $\chi^\prime$, of the 
flow birefringence data\cite{pst2} for the startup flows, via a relation 
similar to Eqn.\ \ref{creep2}, i.e., 
\beq 
\lambda \sim \tan^2 \chi^\prime. 
\label{chiinfty}
\eeq 

\begin{figure}
\centerline{\epsfxsize = 8cm \epsfbox{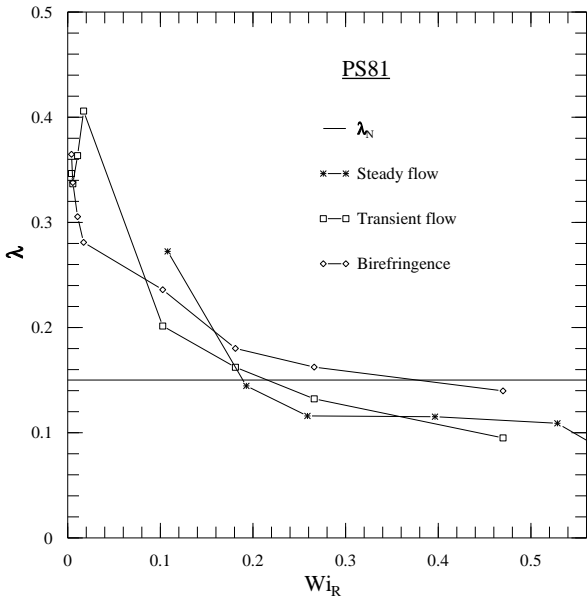}}
\caption{The Weissenberg number ($Wi_R = \dot{\gamma} \lambda^{1/2} \tau_R$) 
dependence of the flow-type parameter, $\lambda$, for the polystyrene sample 
PS81, extracted from three different means (as explained in the text) and 
plotted together along with $\lambda_N = 0.1501$.  The Weissenberg number is 
calculated based on the measured values of $\dot{\gamma}$, $\lambda$ and 
$\tau_R$.}
\label{lstrn81} 
\end{figure}

We have presented the results in Figs.\ \ref{lstrn81}, \ref{lstrn82} and 
\ref{lstrn2} for the samples PS81, PS82, and PS2 respectively and compared 
the results with the theoretical value of $\lambda_N = 0.1501$.  For each 
of these three polymeric samples, the overall qualitative match of the 
data as well as the quantitative match between the $\lambda$ values 
calculated using three different experimental procedures, is quite 
satisfactory.  The fair quantitative agreement between the asymptotic 
values of $\lambda$ obtained from the transient flow experiments and the 
steady flow data for $\lambda$ confirms that the startup experiments with 
the entangled samples have almost reached their steady values in the total 
predetermined evolution time of $t_e = 30$ s\cite{pst2}.  In agreement with 
the steady-state result, the transient experiments also show $\lambda$ values 
exceeding \lN at the lowest rates of deformation studied, which again 
are higher than $Wi_R = 0$, where we should expect that $\lambda = 
\lambda_N$.  In fact, for PS81, at the three lowest rates of 
deformations studied, the extracted value of $\lambda$ from the 
transient experiments do show a trend to decrease after reaching 
a maximum.  

\begin{figure}
\centerline{\epsfxsize = 8cm \epsfbox{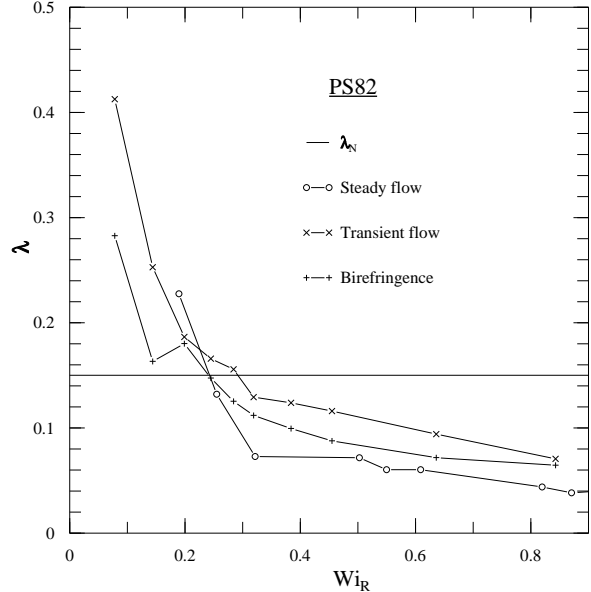}}
\caption{Same as in Fig.~14, but for the polystyrene sample PS82.}
\label{lstrn82} 
\end{figure}

\subsection{The DEMG model comparison of the steady flow TCFB results}

Probably one of the most used theoretical models to describe the dynamics of 
entangled polymers is the Doi-Edwards (DE) tube (reptation) model\cite{DE}.  
The primary drawback of this model, as is clearly evident by its failure 
to predict the polymer dynamics at high deformation rates \gdotsql $> 
1/\tau_R$, is the fact that the primitive chain is assumed to be 
inextensible.  Subsequently, in an effort to improve its predictions, the 
Doi-Edwards-Marrucci-Grizzuti (DEMG) model\cite{M&G,mead1} was developed to 
incorporate chain-stretching into the original DE tube model.  As noted 
earlier, the DEMG model contains two widely separated time scales, one for 
the relaxation of orientation ($\tau_d$) and a much shorter one for the 
relaxation of the stretch of polymer segments ($\tau_R$).  Thus, as the 
strain-rate \gdotsql is increased, segmental orientation takes place 
first when \gdotsql $\sim 1 / \tau_d$ or $Wi_d = $ \Wid $= {\cal O}(1)$ 
and segmental stretching then ``starts'' at a much larger strain-rate 
when $Wi_R = $ \WiR $= {\cal O}(1)$.  Also, it is expected 
that segmental orientation will take place earlier in dilute systems than 
in entangled systems.  In agreement with the aforesaid expectations, a 
number of numerical studies\cite{pearson2} have suggested that the DEMG 
model has more success over the DE model when compared to experimental 
results for startup of a simple shear flow at high deformation rates 
($Wi_R \geq 1$).  Following this, in a recent numerical 
study\cite{mead1,mead2}, the DEMG model was developed further by ``adding'' 
a non-linear finitely extensible spring force. 

\begin{figure}
\centerline{\epsfxsize = 8cm \epsfbox{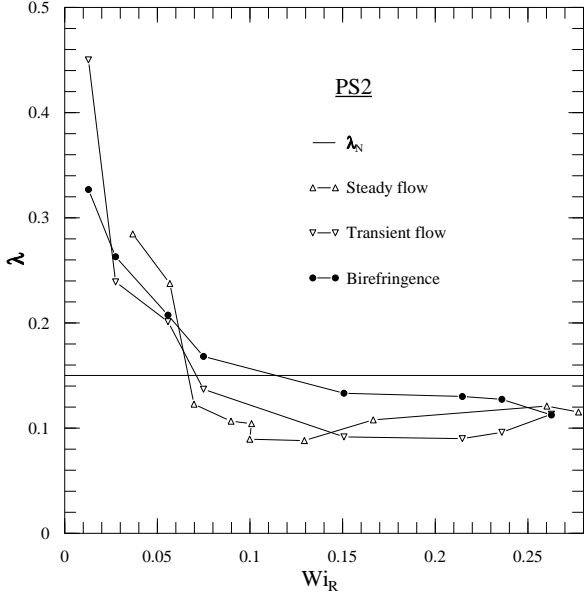}}
\caption{Same as in Fig.~15, but for the polystyrene sample PS2.}
\label{lstrn2} 
\end{figure}

\begin{figure}
\centerline{\epsfxsize = 8cm \epsfbox{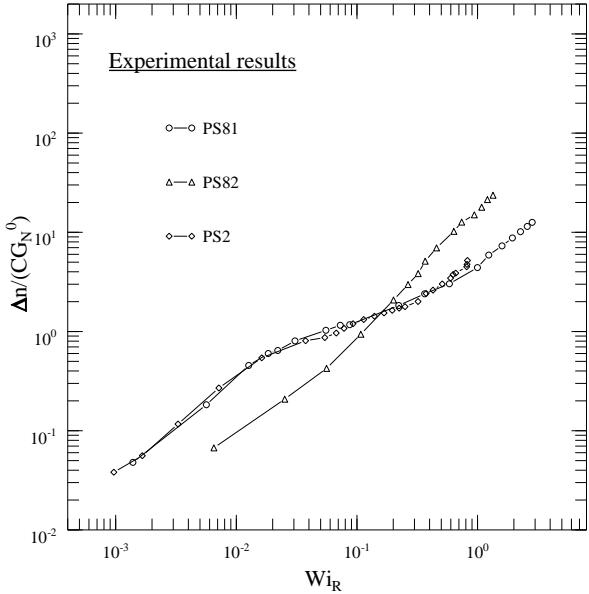}}
\caption{The experimental steady-state birefringence data for the three 
entangled polystyrene solution plotted as functions of the Weissenberg 
number, $Wi_R = \dot{\gamma} \lambda^{1/2} \tau_R$, based on the measured 
magnitude of the principal eigenvalue, $\dot{\gamma} \lambda^{1/2}$, of 
the velocity-gradient tensor, ${\bf \nabla \vec{v}}$, and the Rouse time 
scale, $\tau_R$.  The birefringence is normalized with $C G_N^0$, where 
$G_N^0$ is the plateau modulus, and $C$ is the stress-optical coefficient.  
The data are plotted in the same scale as the corresponding DEMG model 
predictions in Fig.~18.}
\label{expn}
\end{figure}

\begin{figure}
\centerline{\epsfxsize = 8cm \epsfbox{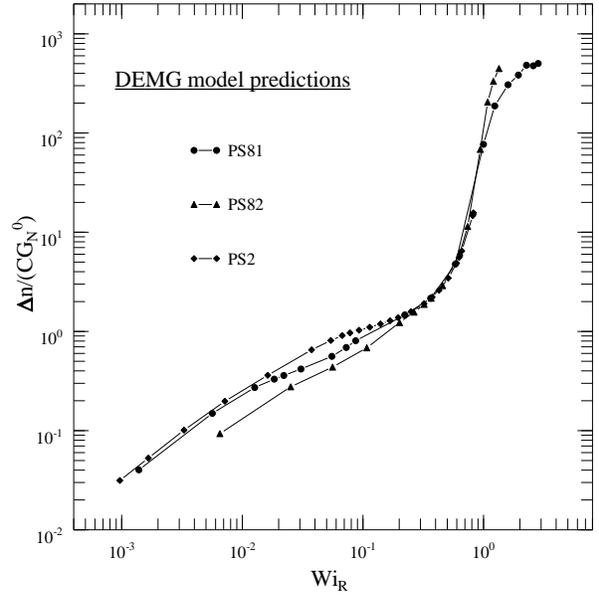}}
\caption{The DEMG model predictions for the steady-state $\Delta n/(C G_N^0)$, 
using measured values of both $\dot{\gamma}$ and $\lambda$, at several 
different strain-rates $Wi_R$ but the same set of rollers, $\lambda_N = 
0.1501$, for the polystyrene samples.  The model parameters $n_t$, $N_e$, 
and $\tau_R$ are given in Table~I.}
\label{demgn}
\end{figure}

For an entangled solution, birefringence measurements directly relate to 
changes in the conformation of the polymer molecules.  As mentioned earlier, 
the complete characterization of the flow-induced anisotropy of a polymeric 
fluid requires determination of the birefringence, $\Delta n$, as well as 
the orientation angle, $\chi$, of the principal axes of the refractive 
index tensor.  In this section, we present the steady-state two-color 
flow birefringence results and compare them with the predictions of the 
DEMG model using the flow parameters measured via DLS as input.  We 
assume that the velocity-gradient tensor in $\sim 75 \mu$m diameter 
birefringence measurement zone surrounding the stagnation point of the 
co-rotating two-roll mill can be approximated by Eqn.\ (\ref{nablav}) and 
(\ref{symm}).  The values of the model parameters $N_e$, $\tau_R$, and $n_t$ 
calculated for the three solutions are given in the Table\ \ref{samples}.  
Here $n_t$ is the number of Kuhn statistical subunits in the chain based on 
the assumption of 10 monomers per subunit and is calculated from the molecular 
weight of the polymer\cite{flory}.  The steady-state experimental results 
for the dimensionless birefringence, $\Delta n/(C G_N^0)$, for the three 
entangled samples are presented in Fig.\ \ref{expn} as functions of 
$Wi_R$.  The birefringence, $\Delta n$, is scaled by the birefringence 
$C G_N^0$ (obtained by using the stress-optical law) that would be 
present at a stress level equal to the plateau modulus, $G_N^0$.  
Here, $C$ is the stress-optical coefficient which in the free-jointed 
chain model is defined as 
\beq 
C = \frac{2\pi}{45} \left[\frac{(\bar{n} + 2)^2}{\bar{n}}\right] 
\left<\alpha_\parallel - \alpha_\perp\right>, 
\label{C}
\eeq
where $\alpha_\parallel$ and $\alpha_\perp$ are the components of the 
intrinsic molecular polarizability tensor $\underline{\underline{\bf 
\alpha}}(t)$ along and 
perpendicular to the Kuhn statistical segment, and $\bar{n}$ is the bulk 
refractive-index of the medium.  This is justified only up to a modest 
level ($\leq 50 \%$) of fractional chain-extension, so that non-Gaussian 
effects can be neglected.  To scale the experimental birefringence, an 
approximate expression given in Eqn.\ (\ref{G_N}) for the plateau modulus, 
$G_N^0$, and the literature value\cite{J-K} of $C = 5 \times 10^{-10}$ 
cm$^2$/dyn for polystyrene solutions is used.  The model predictions 
corresponding to the experimental curves in Fig.\ \ref{expn} are displayed 
in Fig.\ \ref{demgn} (both on the same scale).  Each 
individual point on the predicted curves of Fig.\ \ref{demgn} is the 
steady-state value obtained from separate runs of DEMG model numerical 
simulations using the measured values of \gdot and $\lambda$ for each 
corresponding point on the experimental curves (Fig.\ \ref{expn}).  The 
model clearly predicts\cite{mead2} the existence of three flow regimes in 
the steady-state, namely, the {\em linear viscoelastic} regime seen at 
low rates of deformation $Wi_d < {\cal O}(1)$, the {\em non-linear 
viscoelastic} regime seen at the intermediate deformation rates lying 
between $Wi_d > {\cal O}(1)$ and $Wi_R < {\cal O}(1)$, and the {\em 
highly non-linear viscoelastic} regime seen at high deformation rates 
$Wi_R > {\cal O}(1)$.  In each of these flow regimes, the effect of 
three parameters, namely, $n_t$, $N_e$, and $\lambda$ on the nature of 
deformation rate dependence of the birefringence, orientation angle, 
stretch and viscosity, as predicted by the DEMG model, has been 
discussed in detail in a recent publication by Mead {\em et al.}\cite{mead2}.  

\begin{figure}
\centerline{\epsfxsize = 8cm \epsfbox{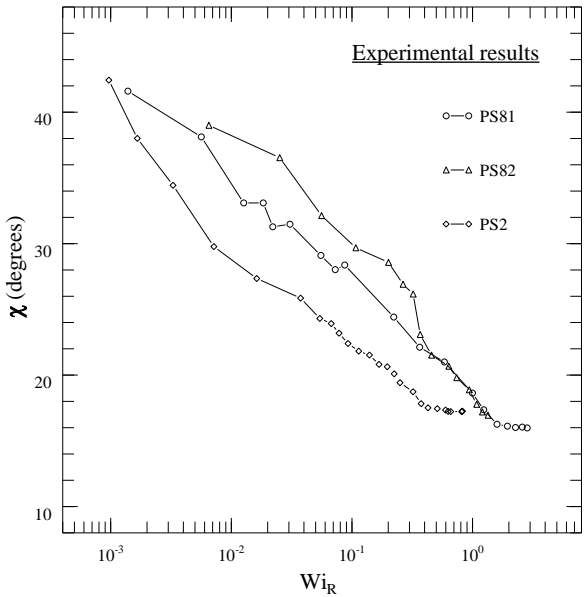}}
\caption{The experimental steady-state orientation angle, $\chi$ (in 
degrees), for the three entangled polystyrene solution plotted as a 
function of the Weissenberg number, $Wi_R = \dot{\gamma} \lambda^{1/2} 
\tau_R$, using the measured values of $\dot{\gamma}$, $\lambda$ and 
$\tau_R$.  The data are plotted in the same scale as the corresponding 
DEMG model predictions in Fig.~20.}
\label{expchi}
\end{figure}

\begin{figure}
\centerline{\epsfxsize = 8cm \epsfbox{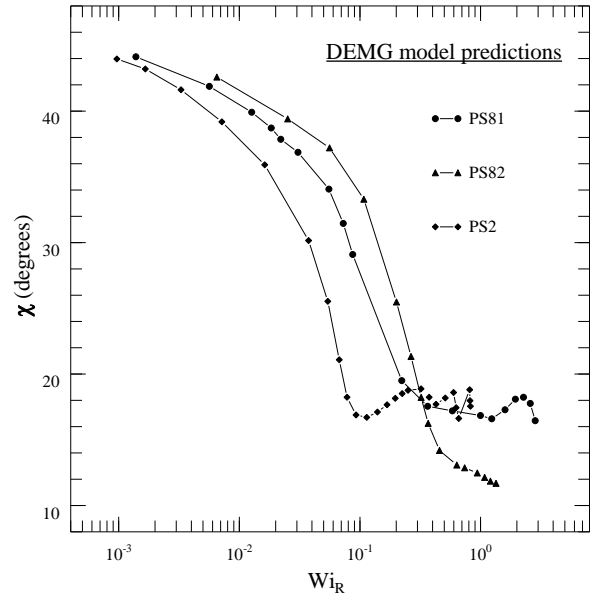}}
\caption{The DEMG model predictions for the steady-state orientation angle, 
$\chi$ (in degrees), using measured values of both $\dot{\gamma}$ 
and $\lambda$, at several different strain-rates, $Wi_R$, for the three 
polystyrene fluids in a two-roll mill with a fixed set of rollers, 
$\lambda_N = 0.1501$.  The model parameters $n_t$, $N_e$, and $\tau_R$ 
are given in Table~I.}
\label{demgchi}
\end{figure}

In the linear viscoelastic regime, the slope of each experimental curve 
on the double-logarithmic plots (Fig.\ \ref{expn}) is indeed 
unity\cite{mead2}, similar to the predicted curves (Fig.\ \ref{demgn}).  
According to the predictions of Ref.\ \cite{mead2}, in this flow regime, the 
relative positions of the birefringence curves from top to bottom for these 
entangled samples 
is primarily determined by the values of $N_e$ and $\lambda$ (curves with 
lower $N_e$ and higher $\lambda$ will fall lower), while $n_t$ has no 
effect.  As clearly seen from Figs.\ \ref{expn} and \ref{demgn}, the 
lower value of $N_e$ for PS82 keeps the $\Delta n/(C G_N^0)$ curve lower 
than those of the other two samples.  The lower value of $\lambda$ for 
PS82 in this regime (Fig.\ \ref{pssteady3}) would predict 
otherwise\cite{mead2}.  The slight difference between the curves for PS2 
and PS81 (both having $N_e \sim 13$) seen both in the experimental and 
predicted plots, is due to the slight difference in the $\lambda$ values 
between the two samples in this regime, as can be clearly seen from Fig.\ 
\ref{pssteady3}.  In this flow regime, although the initial low 
levels of segmental orientation, as predicted in Fig.\ \ref{demgchi}, is 
indeed seen for the experimental curves in Fig.\ \ref{expchi}, the rate 
of decrease of $\chi$ with increasing $Wi_R$, as seen in the experiments 
for all samples, is much faster than predicted.  Also, the orientation 
angles for PS81 at a fixed value of $Wi_R$ in this regime lies in 
between those for the other two liquids which again is in accord with 
the theoretical expectations\cite{mead2} (taking into account the effects 
of the corresponding values of $n_t$, $N_e$ and measured $\lambda$). 

When the second flow regime of non-linear viscoelasticity is approached, 
$Wi_d = {\cal O}(1)$, the scaled birefringence curves are predicted to 
depart from linearity and begin to approach a plateau, as seen in the 
DEMG results of Fig.\ \ref{demgn}.  The width of the plateau is directly 
proportional to $N_e$, and is therefore expected to be about half as wide 
for PS82 ($N_e \sim 7$) as for PS2 and PS81 ($N_e \sim 13$).  The 
experimental curves in this regime do show non-linearity for all three 
samples, but the collapse of the curves and an existence of a narrow 
plateau is seen only for PS81 and PS2 and not for PS82.  The dynamics in 
this regime are controlled by strong segmental orientation of the tube 
without stretching.  Referring to Fig.\ \ref{expchi}, we see that the 
monotonic decrease of the orientation angle with increasing $Wi_R$, seen 
in experiments for all samples is in qualitative agrement with the predictions 
of the DEMG model, resulting from the gradual unwinding and straightening 
out of the tube.  The relative positions of the $\chi$ versus $Wi_R$ 
curves from top to bottom in Fig.\ \ref{demgchi} for the three samples 
are the result of a competition between the effects of $N_e$ and $\lambda$ 
them.  Having a lower value of $\lambda$ (see, Fig.\ \ref{pssteady3}) 
prompts\cite{mead2} the orientational angle for PS82 to fall below those 
for the other two samples, but as can be seen in Figs.\ \ref{expchi} and 
\ref{demgchi}, the effect of $N_e$ is stronger in this case too, since the 
curves with a lower value of $N_e$ (that for PS82, here) are 
predicted\cite{mead2} to show a lower degree of orientation at any fixed 
$Wi_R$ in this flow regime.  Similarly, PS2 exhibits a higher degree of 
orientation than for PS81 at a fixed $Wi_R$, since $\lambda$ is lower 
for PS2 in this intermediate regime of flow.  As $Wi_R \rightarrow 1$, 
the tube is expected to straighten out to its full length and become 
orientated in the direction of the outflow axis [Fig.\ \ref{two-roller1}] 
with the orientation angle reaching its asymptotic value, $\chi^\prime$, 
given in Eqn.\ (\ref{chiinfty}).  Using the asymptotic values of $\lambda \sim 0.1$ for 
PS2 and PS81 and $\lambda \sim 0.05$ for PS82 from Fig.\ \ref{pssteady3}, 
we obtain $\chi^\prime \sim 17.55^\circ$ for PS2 and PS81 and $\chi^\prime 
\sim 12.60^\circ$ for PS82 which are in agrement with the results in Fig.\ 
\ref{demgchi}.  Interestingly, the theoretical predictions for the 
orientation angles for PS2 and PS81 in Fig.\ \ref{demgchi} show an 
``undershoot'' before reaching $\chi^\prime$.  For the highest rates 
of deformation, $Wi_R \sim 1.5$, studied for PS82, the predicted 
orientation angle has not reached its asymptotic value.  The 
experimental curves of $\chi$, as seen in Fig.\ \ref{expchi}, 
qualitatively follow the behavior shown by the predicted curves except 
for that they do not show an undershoot behavior.  In addition, 
$\chi^\prime \sim 17.5^\circ$ and $\chi^\prime \sim 16^\circ$ for PS2 
and PS81, respectively and although the final value of $\chi$ at the 
highest $Wi_R$ studied for PS82 is $\sim 17.4^\circ$, it does not seem 
to have saturated; and for $0.3 < Wi_R \leq 1.5$ the $\chi$ values for 
PS81 and PS82 overlap.

The third flow region of highly non-linear viscoelasticity ($Wi_R \geq 1$) 
is predicted\cite{mead2} to show three distinct signatures.  Firstly, the 
tube orientation should be complete as indicated by $\chi = \chi^\prime$.  
We have discussed about the behavior of the predicted and experimental 
orientation angles for all polystyrene samples in the preceding paragraph.  
Secondly, the onset of chain-stretching is predicted to take place with a clear 
evidence of the dimensionless birefringence exceeding its plateau value 
of unity as shown in Fig.\ \ref{demgn}.  This is evident in the experiments 
too, as displayed in Fig.\ \ref{expn}.  Thirdly, the onset of chain-stretching 
is also marked by the collapse of all the birefringence curves 
to a single universal curve at the end of the plateau i.e., they should 
follow the scaling behavior $\Delta n/(C G_N^0) = f(Wi_R)$ in this region, 
as is clearly shown by the predicted traces in Fig.\ \ref{demgn}.  This 
feature is well demonstrated in the experiments (Fig.\ \ref{expn}) by 
PS81 and PS2 but not by PS82, which fails to show a well-defined plateau 
too.  At high rates of deformation ($Wi_R > 1$), the chain-stretching 
dynamics prompts the different birefringence curves to reach different 
asymptotes proportional to $\sim n_t/N_e$.  The theoretical curves in 
Fig.\ \ref{demgn} show these features: the increase in the birefringence 
with $Wi_R$ in this regime is faster for PS82 ($n_t = 8420$ and $N_e 
\sim 7$) than for PS81 ($n_t = 8420$ and $N_e \sim 13$), which is in 
turn faster than for PS2 ($n_t = 2890$ and $N_e \sim 13$).  However, 
none of the $\Delta n/(C G_N^0)$ curves are predicted to reach their 
asymptotic values ($\sim 648$ for PS81, $\sim 1203$ for PS82 and $\sim 222$ for 
PS2) for the highest $Wi_R$ studied, though the curves for PS81 and PS82 
indeed show a tendency towards saturation in Fig.\ \ref{demgn}.  At the 
highest rates of deformations, the experimental birefringence levels 
(Fig.\ \ref{expn}) for all samples are significantly below their saturation 
values, and they do not show a tendency to saturate.  Contrary to the 
prediction, the experimental $\Delta n/(C G_N^0)$ curve for PS81 falls 
below that for PS2.  We note that the relative level of birefringence for 
the different samples as depicted by the individual curves in Figs.\ 
\ref{expn} and \ref{demgn} are strongly dependent on the choice of the 
value of the relaxation modulus, $G_N^0$ [Eqn.\ (\ref{G_N})].  However, 
we believe that the qualitative as well as quantitative agreement between 
the experimental and predicted birefringence curves well justifies the 
choice of value for $G_N^0$.  

Following Mead {et al.}\cite{mead1}, we define the generalized viscosity 
function (which is the conventional viscosity appropriately generalized 
for mixed shear and extensional flow) as follows: 
\beq 
\eta (\dot{\gamma}, \lambda) = \frac{\sigma_{xy}}{\dot{\gamma}(1 + \lambda)}. 
\label{eta'} 
\eeq 
This reduces to the usual definition of the shear viscosity and the planar 
extensional viscosity in the limits of $\lambda = 0$ and $\lambda = 1$, 
respectively.  The scaling factor $(1 + \lambda)$ makes the viscosity 
independent of the flow-type $\lambda$ at low strain-rates.  In the 
principal axes of the rate of deformation tensor, which are at $45^\circ$ 
from the coordinate axes used in Figs.\ \ref{two-roller1} and \ref{two-roller2}, 
we get $\eta (\dot{\gamma}, \lambda) = \frac{\sigma_{xx}^\prime - 
\sigma_{yy}^\prime}{\dot{\gamma} (1 + \lambda)}$.  Using the stress-optical 
relationship\cite{J-K}, which states that the birefringence tensor is 
proportional and coaxial to the stress tensor, one can relate the 
birefringence, which is the difference of the principal values of the 
refractive-index tensor, $\underline{\underline{\bf n}}$, in the plane of 
the flow (i.e., $\Delta n = n_{xx}^\prime - n_{yy}^\prime$), to the stress 
as $\sigma_{xy} = \frac{\Delta n}{2C}\sin(2\chi)$, so that Eqn.\ (\ref{eta'}) 
can be written as 
\beq 
\eta (\dot{\gamma}, \lambda) = \frac{\Delta n \sin(2\chi)}{2C \dot{\gamma} 
(1 + \lambda)}.
\label{eta} 
\eeq 

\begin{figure}
\centerline{\epsfxsize = 8cm \epsfbox{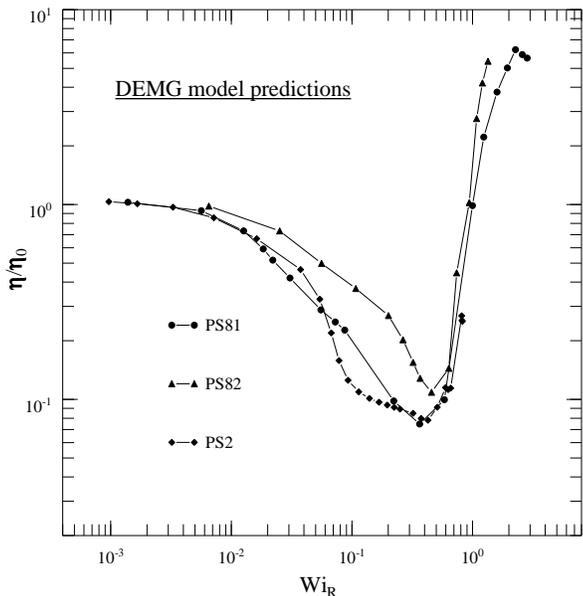}}
\caption{The predicted behavior of the steady-state generalized extensional 
viscosity scaled with the zero-shear viscosity, as a function of $Wi_R$ for 
the polystyrene solutions, obtained by using the computed values of $\Delta n$ 
and $\chi$ from the DEMG model, the measured values of $\dot{\gamma}$ and 
$\lambda$, the stress-optical relationship, and the definitions of $\eta$ 
and $\eta_0$.  The model parameters $n_t$, $N_e$, and $\tau_R$ are given in 
Table~I.}
\label{demgeta}
\end{figure}

We present the predicted and the experimental steady-state viscosity 
function versus the measured $Wi_R$ for the three polystyrene solutions 
in Figs.\ \ref{demgeta} and \ref{expeta}, respectively.  The predicted 
values of $\eta$ are calculated from Eqn.\ (\ref{eta}) by using the 
predicted birefringence and extinction angle shown in Figs.\ \ref{demgn} 
and \ref{demgchi}, and by scaling the viscosity with the zero shear 
viscosity, 
\beq 
\eta_0 = \frac{G_N^0 \tau_d \pi^2}{45}.
\label{demgeta0} 
\eeq 
Similarly the experimental plots shown in Fig.\ \ref{expeta} are calculated 
with the use of the measured $\Delta n$ (Fig.\ \ref{expn}) and $\chi$ 
(Fig.\ \ref{expchi}) and then by normalizing by the experimental $\eta_0$ 
given in Table\ \ref{samples}.  In sharp contrast to the 
observation\cite{sridhar} that dilute solutions, (e.g., Boger fluids) show 
strain-rate thickening at high strain-rates, for entangled solutions the 
viscosity function is predicted to show different behaviors in the three 
earlier defined flow regimes.  In the first flow regime of linear 
viscoelasticity, the viscosity is predicted to be nearly constant and equal 
to its zero shear-rate value, $\eta_0$, (Fig.\ \ref{demgeta}).  This is 
approximately followed by the experimental curves (Fig.\ \ref{expeta}).  In 
the intermediate flow regime [between $Wi_d > {\cal O}(1)$ and $Wi_R < 
{\cal O}(1)$], where the stress is generated primarily via orientation of 
the tube segments, the model predicts a strong strain-rate thinning behavior 
with the degree of thinning being an increasing function of $N_e$ and a 
decreasing function of $\lambda$, as shown in Fig.\ \ref{demgeta}.  In the 
experiments, we see that $N_e$ has a stronger effect than $\lambda$ on the 
degree of thinning.  The predicted effect of $\lambda$ on the relative 
positions of $\eta/\eta_0$ versus $Wi_R$ curves for PS81 and PS2 from top 
to bottom is not supported in the experiments. 

\begin{figure}
\centerline{\epsfxsize = 8cm \epsfbox{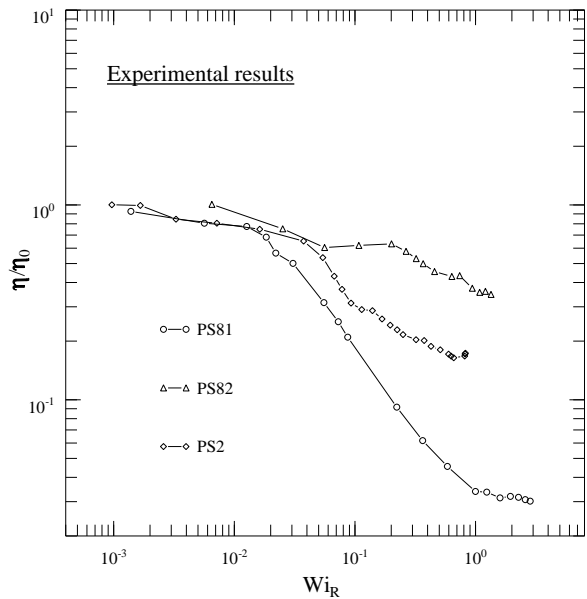}}
\caption{The steady-state generalized extensional viscosity scaled with the 
zero-shear viscosity, versus $Wi_R$ for the polystyrene solutions, obtained by 
using the experimentally measured values of $\Delta n$, $\chi$, $\dot{\gamma}$, 
$\lambda$, $\eta_0$, the stress-optical relationship, and the definition of 
$\eta$.  The data are plotted in the same scale as the corresponding DEMG 
model predictions in Fig.~21.}
\label{expeta}
\end{figure}

The presence of an ``undershoot'' in the predicted orientation angle for PS2 
(Fig.\ \ref{demgchi}) in this flow regime clearly manifests itself in the 
predicted viscosity plot (Fig.\ \ref{demgeta}) too.  This does not seem to 
be the case with PS82, because for PS82 the small undershoot present in the 
predicted orientation angle does not show up in the predicted viscosity.  
Although, the experimental viscosity function shows a thinning behavior with 
the degree of thinning being much weaker in PS82 than the other two solutions, 
the model predicts much stronger viscosity thinning behavior than observed 
in the experiments (except for PS81 where they are similar).  Also, the expected 
``undershoot'' in the viscosity for PS2 is absent in the experiment. 

In the third regime of flow [$Wi_R = {\cal O}(1)$] the model predicts 
a sharp upturn in the viscosity function as chain-stretching is predicted 
to take place.  As the rate of deformation is increased, the viscosity is 
also expected to increase and finally reach an asymptote for $Wi_R >> 1$.  
Surprisingly, however, except for a small upturn shown by the last three 
data points for PS2 in Fig.\ \ref{expeta}, these predicted features are 
completely absent in the experimental data.  The cause for this lies in the 
fact that even though the experimental birefringence values of these samples 
(Fig.\ \ref{expn}) rise above the plateau value, indicating chain-stretching, 
the rise is much weaker than predicted (Fig.\ \ref{demgn}).  Also, the 
degree of orientation of the tube segments in the experiments (Fig.\ 
\ref{expchi}) fails to show the sharpness predicted by the model (Fig.\ 
\ref{demgchi}) in this flow regime.  

\begin{figure}
\centerline{\epsfxsize = 8cm \epsfbox{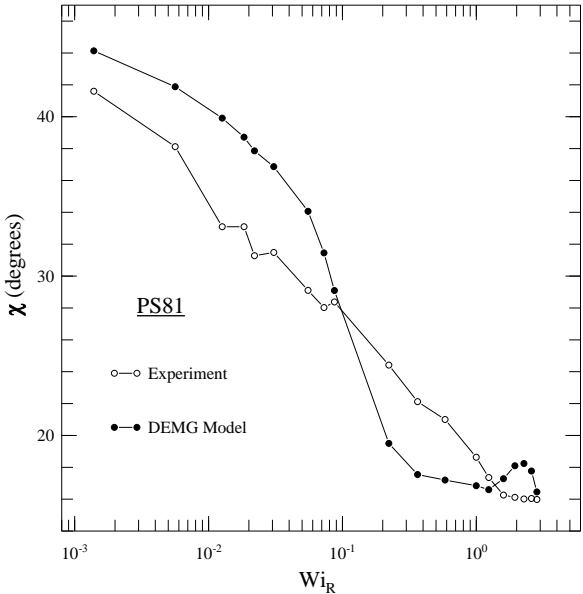}}
\caption{A comparison of the Weissenberg number dependence of the measured 
steady-state orientational angle, $\chi$ (in degrees), with values predicted 
via the DEMG model using the measured values of both $\dot{\gamma}$ and $\lambda$ 
for the PS81 solution.  The model parameters were $N_e = 13$, $n_t = 8420$, 
and $\tau_R = 2.25$~s.}
\label{chi81}
\end{figure}

The above two effects compete with each other to nullify any 
signature of chain-stretching that could have been otherwise seen in the 
generalized viscosity function.  We have noted before that the present 
model\cite{mead1,mead2} has improved the original DEMG model by using 
finitely extensible freely jointed chains instead of infinitely extensible 
Gaussian chains.  In the limit of the high value of $n_t$ (or $M_w$), the 
original version of the model is retrieved and at high deformation rates 
the chains stretch tremendously to a show a nearly singular viscosity or 
birefringence.  This effect is apparent in the model 
predictions shown in Figs.\ \ref{demgeta} and \ref{demgn}, where both PS81 
and PS82 (higher $n_t$ or $M_w$) show a steeper increase in $\eta$ and 
$\Delta n/(C G_N^0)$ at $Wi_R \sim 1$) compared to PS2, but is absent in the 
experiments (Figs.\ \ref{expeta} and \ref{expn}).  Unfortunately, 
the highest dimensionless rate of deformation or $Wi_R$ that we could 
reach in the experiments is only about 3 for PS81 and about 1 for the 
other two solutions.  This deficiency is caused by the fact that the 
viscoelastic modification relative to the Newtonian flow was much stronger 
than we had initially expected.  The measured values of \gdot and $\lambda$ 
were thus reduced compared to their Newtonian values, so that the originally 
calculated theoretical $(Wi_R)^{\mbox{max}}_{N} \equiv (\dot{\gamma}_N 
\sqrt{\lambda}_N \tau_R)^{\mbox{max}} = {\cal O}(10)$ corresponding to the 
maximum motor speed chosen for the experiments ultimately produced only 
$(Wi_R)^{\mbox{max}} \equiv (\dot{\gamma} \sqrt{\lambda} \tau_R)^{\mbox{max}} 
= {\cal O}(1)$.  The absence of chain-stretching signatures in the viscosity 
may also have to do with the choice of sample compositions which gives rise 
to low values for the number of entanglements per chain.  It would be 
interesting to carry out further experiments on samples with higher $N_e$ in 
this highly non-linear viscoelastic flow regime at high enough $Wi_R$, 
particularly to look into the chain-stretching effects much more carefully.

\begin{figure}
\centerline{\epsfxsize = 8cm \epsfbox{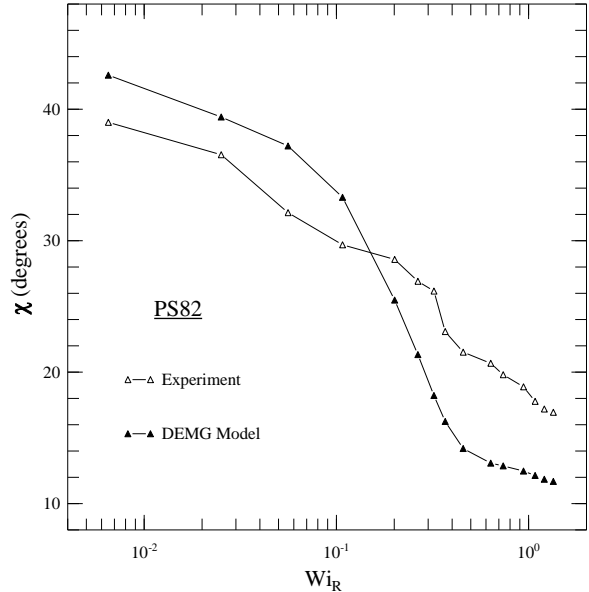}}
\caption{The experimental data for the steady-state, $\chi$ (in degrees), versus 
$Wi_R$ for the sample PS82 compared with the corresponding DEMG predictions, 
obtained by using the measured values of both $\dot{\gamma}$ and $\lambda$, and 
the model parameters $N_e = 7$, $n_t = 8420$, and $\tau_R = 3.01$~s.}
\label{chi82}
\end{figure}

Let us now turn our attention into more quantitative comparisons between 
the steady-state results and the predictions of the DEMG model.  We will 
first consider the case of the orientation angle $\chi$ for three samples 
as shown in Figs.\ \ref{chi81}, \ref{chi82}, and \ref{chi2}.  Since there 
is no other parameter involved for scaling (e.g., in the case of the 
birefringence), direct comparisons between the experimental data and 
model predictions are possible in this case.  Although the experimental 
data follows a similar qualitative decreasing trend in the orientation 
angle with increasing $Wi_R$, as predicted, the quantitative mismatch 
with the prediction is quite obvious in these figures.  In the limit 
$Wi_R \rightarrow 0$, it is expected that $\chi$ should be $45^\circ$, 
corresponding to alignment of the principal optical axis with the 
principal axis of the rate of strain tensor.  The experimental accuracy 
of both $\Delta n$ and $\chi$ at the lowest rates of deformation is 
limited by the sensitivity of the apparatus in determining a minimum 
birefringence and its associated orientation.  Also, the contribution 
from the residual glass birefringence is important at these low levels 
of polymer anisotropy.  At low and intermediate $Wi_R$ the DEMG model 
underpredicts the orientation but at higher $Wi_R$ it overpredicts 
the same, and there is a crossing point between these two behaviors.  
By comparing the results on the orientation angle with those obtained 
from the DEMG model numerical simulations with identical parameters 
but with a constant 
flow-type parameter $\lambda = \lambda_N = 0.1501$, we have confirmed 
that the opposite curvatures seen in the predicted curves below and 
above the crossing point depends on the way $\lambda$ changes with 
$Wi_R$.  The fact that the DEMG model overpredicts the tendency of the 
flow to orient polymer chains towards the outflow axis at intermediate 
and high strain-rates has been observed in earlier experiments\cite{dmitry,jim} 
and it was speculated that ``tube-dilation'', which is not incorporated 
in the DEMG model, may be responsible for such an effect.  In simple 
terms, this means that the effective dilation of the tube radius is 
stronger for large rates of deformation, leading to a reduction of 
$N_e$ and hence the lower orientation angle seen in the experiments.  
Weather or not tube-dilation is present in the system, for $Wi_R 
\geq {\cal O}(1)$, we expect that the polymer molecules will become 
oriented along the outflow axis of the flow-field.  As we have noted 
earlier, and as can be seen clearly from these figures, the asymptotic 
value of $\chi$ is approximately reached in experiments for PS81 and 
PS2, but not for PS82. 

\begin{figure}
\centerline{\epsfxsize = 8cm \epsfbox{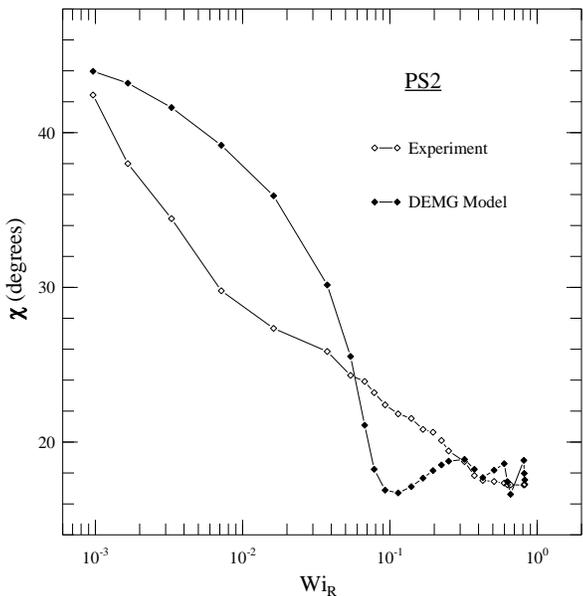}}
\caption{A comparison of the Weissenberg number dependence of the measured 
steady-state orientational angle, $\chi$ (in degrees), with values predicted 
via the DEMG model using the measured values of both $\dot{\gamma}$ and $\lambda$ 
for the PS2 solution.  The model parameters were $N_e = 13$, $n_t = 2890$, 
and $\tau_R = 0.56$~s.}
\label{chi2}
\end{figure}

\begin{figure}
\centerline{\epsfxsize = 8cm \epsfbox{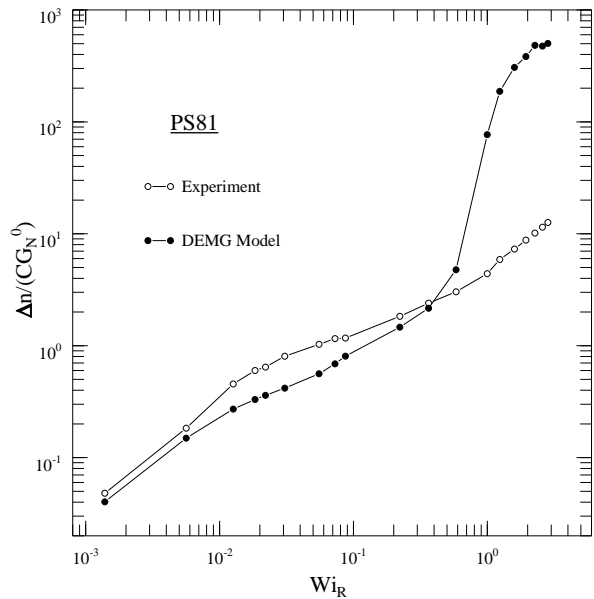}}
\caption{A comparison of the measured steady-state scaled birefringence, 
$\Delta n/(C G_N^0)$ at several different $Wi_R$, with the values predicted 
via the DEMG model obtained by using the measured $\dot{\gamma}$ and $\lambda$ 
for the PS81 solution.  The model parameters used were $N_e = 13$, $n_t = 8420$, 
and $\tau_R = 2.25$~s.}
\label{n81}
\end{figure}

\begin{figure}
\centerline{\epsfxsize = 8cm \epsfbox{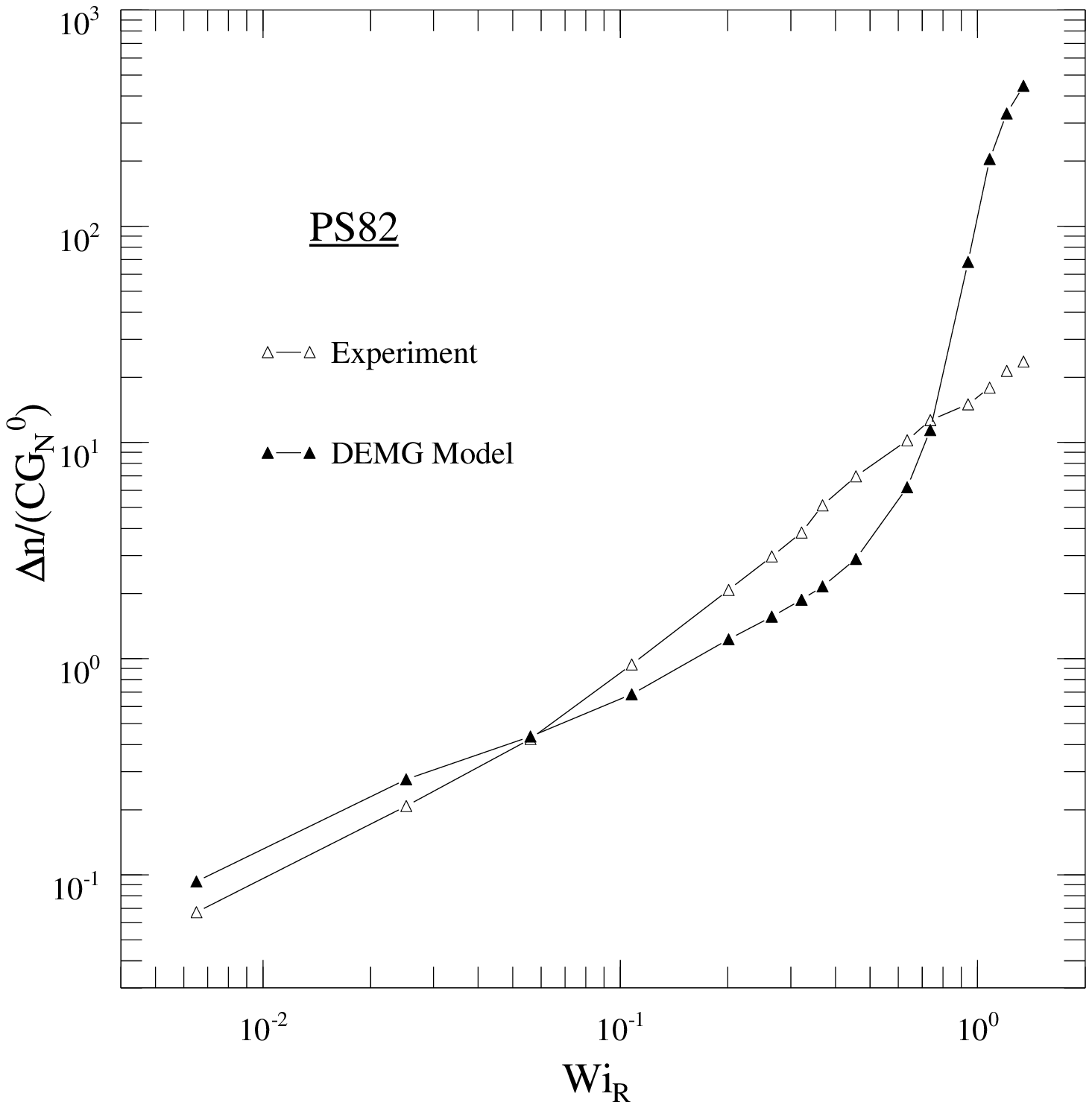}}
\caption{The experimental data for the steady-state $\Delta n/(C G_N^0)$ 
versus $Wi_R$ for the sample PS82 compared with the corresponding DEMG 
predictions, obtained by using the measured values of both $\dot{\gamma}$ and 
$\lambda$, and the model parameters $N_e = 7$, $n_t = 8420$, and $\tau_R 
= 3.01$~s.}
\label{n82}
\end{figure}

The qualitative as well as quantitative match between the experimental 
and predicted birefringence curves for these entangled samples, shown 
in Figs.\ \ref{n81}, \ref{n82} and \ref{n2}, is quite satisfactory.  This 
firstly points to the fact that the choice of the plateau modulus, as 
described earlier in this section, works very well for these entangled 
systems.  Given the proper values of $G_N^0$, we can clearly see that 
DEMG model predictions for birefringence are fairly close to the 
experimental measurements, at least for low and intermediate $Wi_R$.  
We note that the plateau expected in the transition region between 
the dynamics dominated by segmental orientation and segmental stretch, 
is quite narrow, because of the small separation between $\tau_d$ and 
$\tau_R$ ($\tau_d/\tau_R = 3N_e = 21$ for PS82, and $39$ for PS81 \& PS2).  
The plateau is smeared out in the experimental curves, since tube-dilation 
reduces $N_e$ and hence the separation $3N_e$ between these two time scales.  
This supports the tube-dilation idea.  In the nonlinear viscoelastic 
regime, the chain-stretching predicted by the model is much stronger 
than what has been seen experimentally.  In the limit $Wi_R \rightarrow 
\infty$, the model predicts a saturation of birefringence $\sim n_t/N_e$ 
corresponding to a maximum chain-extension ratio $\sim \sqrt{n_t/N_e}$.  
Comparing the maximum birefringence shown in Figs.\ \ref{n81}, \ref{n82}, 
and \ref{n2} with the above numbers, we see that the maximum chain-extension 
predicted by DEMG model at the highest $Wi_R$ studied are $87.8 \%$, 
$63.2 \%$ and $28.5 \%$ for PS81, PS82 and PS2, respectively but the 
experimentally observed maximum chain-extension for these samples is 
only $\sim 15 \%$. 

\begin{figure}
\centerline{\epsfxsize = 8cm \epsfbox{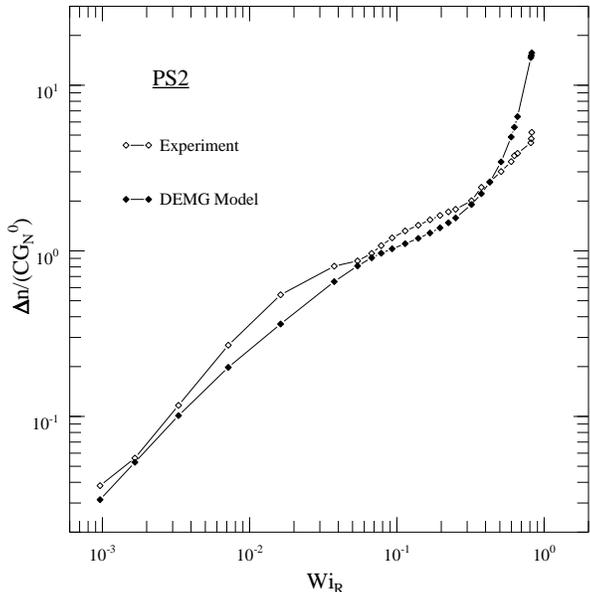}}
\caption{A comparison of the Weissenberg number dependence of the measured 
steady-state birefringence, $\Delta n/(C G_N^0)$,  with the corresponding 
DEMG model predictions obtained by using the measured values of both 
$\dot{\gamma}$ and $\lambda$ for the PS2 solution.  The model parameters were 
$N_e = 13$, $n_t = 2890$, and $\tau_R = 0.56$~s.}
\label{n2}
\end{figure}

The quantitative agreement between the experimental and the predicted 
$\eta/\eta_0$ for small rates of deformation in Figs.\ \ref{eta81}, 
\ref{eta82}, and \ref{eta2} directly points to a proper choice of the 
calculated and the experimental values of $\eta_0$.  In the intermediate 
flow regime the DEMG model predicts a strong strain-rate thinning 
behavior for the generalized extensional viscosity for entangled 
solutions.  This also distinguishes the entangled solution from dilute 
ones, for which a monotonic increase in $\eta$ with the increase of \gdot 
is expected.  In the experiments, we see that the solutions with $N_e 
\sim 13$ show much stronger thinning than the solutions with a lower 
value of $N_e$, consistent with the prediction.  It has been shown in 
earlier experiments that the original DE model overpredicts the strength 
of shear-thinning in simple shear flows\cite{jim} and the DEMG model also 
shows a similar behavior in extension-dominated flows\cite{dmitry}.  
Similarly, we note that the excessive shear-thinning predicted by DEMG 
model for PS82 and PS2 (shown in Figs.\ \ref{eta82} and \ref{eta2}, 
respectively) is due to the fact that the predicted orientation angle 
and the birefringence are too small in this intermediate range of $Wi_R$ 
(see, Figs.\ \ref{chi82}, \ref{chi2}, \ref{n82} and \ref{n2}).  
The tendency of DEMG model to over-orient the polymer chain away from 
the principle axis of the rate of strain tensor, is 
speculated\cite{dmitry,jim} to be due to the fact that the model does 
not incorporate the tube-dilation effect which may be present in the 
real systems.  For PS81, as can be seen from Fig.\ \ref{eta81}, the 
experimental curve for the viscosity closely follows the predictions 
of the DEMG model in the low and intermediate range of $Wi_R$ primarily 
due to the fact that, in this region, $\Delta n$ is too small but 
$\chi$ is too large so as to compensate each other.  It can be seen 
from these figures that as $Wi_R$ approaches ${\cal O}(1)$, the model 
predicts an upturn of the viscosity followed by a sharp increase owing 
to the prediction of a strong chain-stretching phenomenon.  Apart from 
a small upturn seen in Fig.\ \ref{eta2},  the model fails to account 
for the observed behavior of the generalized extensional viscosity in 
this non-linear viscoelastic regime.

\subsubsection{Cox-Merz superposition} 

The empirical Cox-Merz relation states that the steady shear viscosity 
is equal to the modulus of the complex dynamic viscosity evaluated at 
the the angular frequency equal to the shear-rate.  This is observed 
to be valid mostly in the linear viscoelastic regime\cite{wisbrun}, 
and also in the case of a simple shear flow ($\lambda = 0$), primarily 
because chain-stretching is not significant in either of these cases.  
Thus, for extensional flows ($\lambda > 0$) departure from Cox-Merz 
``rule'' constitutes another means to study chain-stretching effects 
present in a given flow system.  Following a suggestion of Mead {\em et 
al.} in Ref.~\cite{mead2}, we have compared the measured generalized 
extensional viscosity, $\eta$, for the steady-state flow generated 
in a two-roll mill as a function of the measured velocity-gradient 
(or ``shear-rate'') \gdot for the entangled solutions with the 
corresponding dynamic linear viscoelastic measurement, i.e., 
$|\eta^\star(\omega)|$, in the form of ``Cox-Merz plots'' shown in 
Figs.\ \ref{eta81}, \ref{eta82}, and \ref{eta2}.  Both parameters are 
scaled with the zero-shear viscosity $\eta_0$ (see, Table\ \ref{samples}).  
The match between the two sets of data at low deformation rates for all 
samples and that the values for these parameters are in the same ballpark 
provides an increased confidence on the choice of the corresponding 
$\eta_0$ values.  At low and intermediate range of flow deformations, 
the values of these two sets of parameters are in a closer agreement for 
the samples with $N_e \sim 13$ than that for PS82 ($N_e \sim 7$).  Apart 
from this, the overall dissimilarities between the deformation rate 
dependence of these two sets of results is obvious from these plots.  
At high deformation rates, chain-stretching effects become important, 
and, though it is not as strong as demanded by the DEMG model seen in 
Figs.\ \ref{eta81}, \ref{eta82}, and \ref{eta2} in the previous section, 
its effect is strong enough to show an increased departure of $\eta$ 
from the Cox-Merz superposition.  This departure is quite expected, as 
we have noted above, since the chain-stretching effects are not accounted 
for in this empirical rule.  The departure seen in the intermediate range 
of deformation rates is because of the fact that even in this regime, 
the flow experienced by the polymers in the stagnation region of a 
two-roll mill is very different from a shear-flow.  In short, these 
figures provide a quantitative picture of the relative importance of 
chain-stretching to orientation effects. 

\begin{figure}
\centerline{\epsfxsize = 8cm \epsfbox{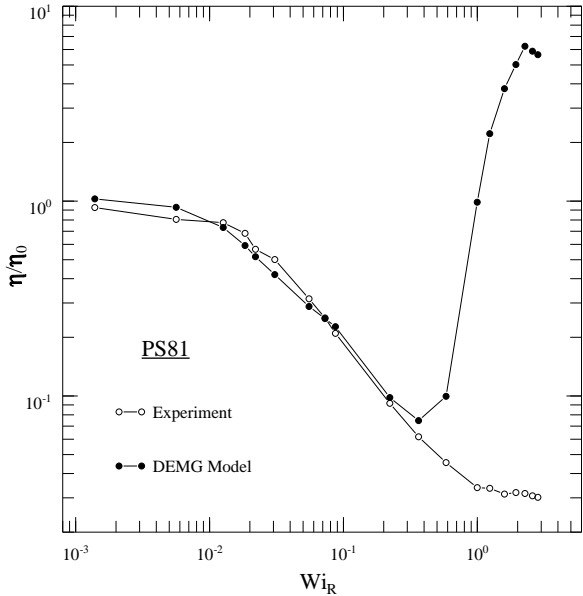}}
\caption{The measured steady-state generalized extensional viscosity for the 
PS81 solution scaled with the zero-shear viscosity at several different $Wi_R$ 
is compared with that computed from the DEMG model.  The experimental results 
were obtained from the measured values of $\Delta n$, $\chi$, $\dot{\gamma}$, 
$\lambda$, $\eta_0$, and using the stress-optical relationship and the 
definition of $\eta$.  The theoretical values were obtained from the computed 
values of $\Delta n$, $\chi$ and $\eta_0$ using the DEMG model with the measured 
values of $\dot{\gamma}$ and $\lambda$ as input.  The model parameters were 
$N_e = 13$, $n_t = 8420$, and $\tau_R = 2.25$~s.}
\label{eta81}
\end{figure}

\begin{figure}
\centerline{\epsfxsize = 8cm \epsfbox{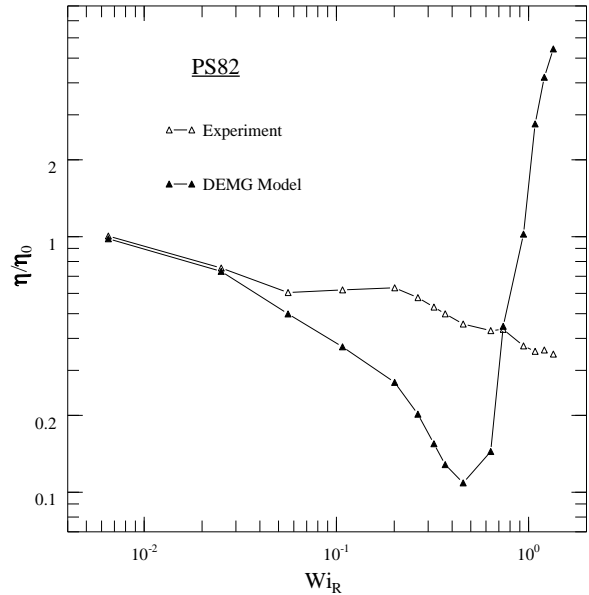}}
\caption{A comparison of the experimental steady-state scaled generalized 
extensional viscosity $\eta/\eta_0$ versus $Wi_R$ for the PS82 fluid 
(obtained from the measured values of $\Delta n$, $\chi$, $\dot{\gamma}$, 
$\lambda$, $\eta_0$, and with the use of the stress-optical relationship and 
the definition of $\eta$), with the predicted values obtained from the computed 
$\Delta n$, $\chi$ and $\eta_0$, by using the DEMG model ($N_e = 7$, $n_t = 
8420$, and $\tau_R = 3.01$~s.) with the measured $\dot{\gamma}$ and $\lambda$ 
as input.}
\label{eta82}
\end{figure}

\section{Summary and Conclusions} 

The recently developed\cite{wang} dynamic light scattering section of  
the two-color flow birefringence experimental setup is used to 
characterize the steady-state flow-fields for a Newtonian fluid as well 
as for viscoelastic, entangled polymeric fluids in a two-roll mill, by 
measuring the velocity-gradient and the flow-type parameter for these 
fluids.  Assuming Newtonian, creeping flow symmetry of the flow-field, 
the flow parameters for a Newtonian fluid, obtained directly from the 
decay rate of the autocorrelation functions measured at the stagnation 
region of the flow, are shown to be very well described by the Frazer's 
creeping flow solution\cite{dunlap}. 

The measured flow parameters for the polymeric fluids show clear 
departures from their corresponding Newtonian values, providing direct 
evidence of flow modification due to the conformational dynamics of the 
polymer molecules.  Our results suggest that this departure may have a 
stronger dependence on the entanglement density of the polymer chains 
than on the molecular weight and concentration. 

\begin{figure}
\centerline{\epsfxsize = 8cm \epsfbox{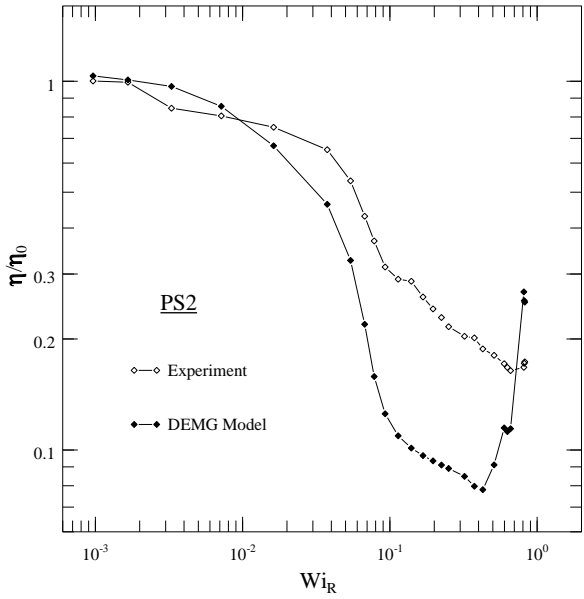}}
\caption{A comparison of the steady-state generalized extensional viscosity 
for the PS2 solution scaled with zero-shear viscosity versus $Wi_R$, and that 
computed from the DEMG model.  The experimental results were obtained from the 
measured values of $\Delta n$, $\chi$, $\dot{\gamma}$, $\lambda$, $\eta_0$, and 
by using the stress-optical relationship and the definition of $\eta$.  The 
theoretical values were obtained from the computed values of $\Delta n$, $\chi$ 
and $\eta_0$, by using the DEMG model with the measured $\dot{\gamma}$ and 
$\lambda$ as input.  The model parameters used were $N_e = 13$, $n_t = 2890$, 
and $\tau_R = 0.56$~s.}
\label{eta2}
\end{figure}

\begin{figure}
\centerline{\epsfxsize = 8cm \epsfbox{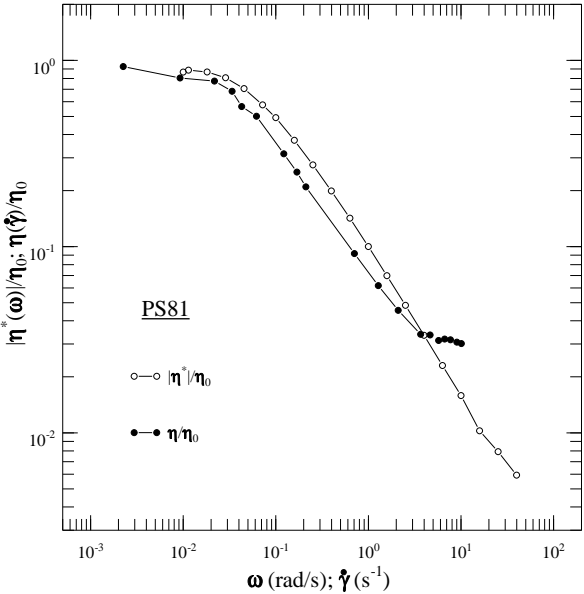}}
\caption{The modulus of the complex viscosity, $|\eta^\star|$, versus the 
angular frequency, $\omega$ (in rad/s), for the polystyrene solution PS81 is 
compared with the steady-state generalized extensional viscosity, $\eta$, 
versus the measured velocity-gradient, $\dot{\gamma}$ (in s$^{-1}$): the 
``Cox-Merz plot''.  Both results are scaled by the measured zero-shear 
viscosity.}
\label{coxm81}
\end{figure}

\begin{figure}
\centerline{\epsfxsize = 8cm \epsfbox{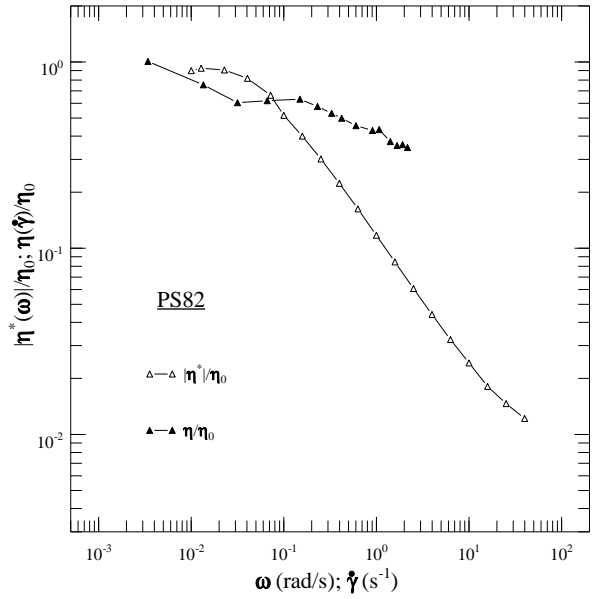}}
\caption{A comparison of $|\eta^\star|$ and $\eta$ for the sample PS82 
plotted against the radial frequency, $\omega$ (in rad/s), and the measured 
velocity-gradient, $\dot{\gamma}$ (in s$^{-1}$), respectively: the 
``Cox-Merz plot''.  Both results are scaled by the measured zero-shear 
viscosity.}
\label{coxm82}
\end{figure}

\begin{figure}
\centerline{\epsfxsize = 8cm \epsfbox{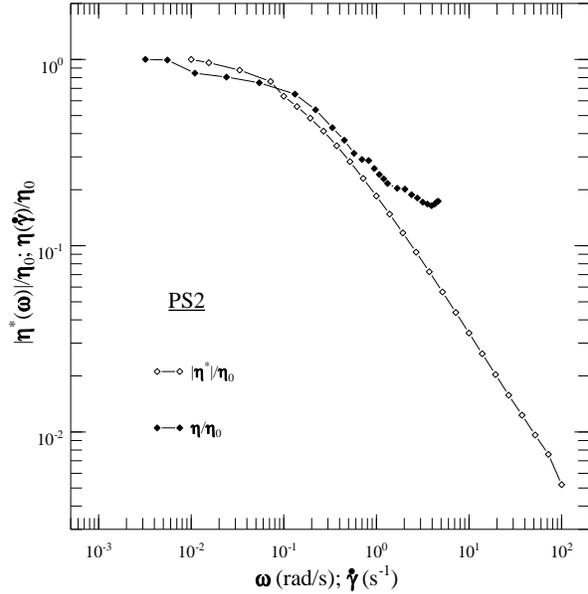}}
\caption{The ``Cox-Merz plot'' for the polystyrene PS2 fluid, i.e., a 
combined plot of the scaled modulus of the complex viscosity, 
$|\eta^\star|/\eta_0$, versus the angular frequency, $\omega$ (in rad/s), 
and the scaled steady-state generalized extensional viscosity, $\eta/\eta_0$, 
versus the measured velocity-gradient, $\dot{\gamma}$ (in s$^{-1}$).}
\label{coxm2}
\end{figure}

Within the limit of experimental error, the flow-field at the stagnation 
region of the two-roller is verified to retain its symmetry, for all rates 
of deformation studied with the Newtonian fluid and for the strain-rates 
\gdotsqlN $< 0.5$ \si ($Wi_R < 1$) with the polystyrene samples.  The 
flow-type parameter for the polystyrene fluids, extracted from the steady-state 
DLS experiments, was found to exceed its Newtonian value at very low rates 
of deformation.  This result was verified to be consistent from three 
different experimental means.  

The dynamics of the entangled polymers under study was coupled with 
the changes in the flow-field.  As we have noted above, the DLS 
experiments has probed the effect of the viscoelasticity on the velocity 
field compared to its Newtonian form.  On the other hand, the polymer 
response induced by the flow-fields generated by a two-roll mill is 
studied using two-color flow birefringence experiments, which were in turn 
compared with the birefringence predictions from the DEMG model using 
the measured flow data as input to the model. 

Similar to the model predictions, the experimental birefringence 
results clearly illustrate the existence of three steady-state 
flow regimes.  In the first two regimes of low and moderate 
rates of deformation, where the dynamics is dominated by chain 
segment re-orientation and reptational diffusion, the model 
predictions are qualitatively, and some times, quantitatively, 
reproduced in the experiments.  However, the model is clearly inadequate 
in describing the polymer dynamics at sufficiently high rates of 
deformation, where chain-stretching is important.  Our data do 
show signatures of chain-stretching: the experimental birefringence  
exceeds the plateau value, the Weissenberg number scaling behavior 
shown at the onset of chain-stretching by at least two of the three 
polymer samples studied, and the departure of the generalized extensional 
viscosity data from the empirical ``Cox-Merz superposition''.  The 
comparisons of the 
experimental birefringence as well as the generalized extensional 
viscosity data unambiguously indicates that the DEMG model 
overpredicts chain-stretching in these entangled samples.  The 
relative values of birefringence, orientation angle, and viscosity 
for different entangled polymeric samples are shown to be determined 
by the competition between the effects of the corresponding flow-type 
parameter, molecular weight (via $n_t$) and the number entanglements 
per chain.  Our experiments have demonstrated that the effect of $N_e$ on 
these parameters is much stronger than that of the $M_w$ and $\lambda$.  
We note that the model fails in describing the ``smearing'' of the 
expected plateau region in the birefringence curves.  In particular, 
it overpredicts the tendency of the flow to rotate the polymer chains 
toward the outflow axis, at high Weissenberg numbers, 
thereby predicting excessive thinning of the generalized viscosity 
in this regime of flow for at least two of the polystyrene samples 
of our study.  It was suggested in an earlier paper\cite{dmitry} that 
one of the prime causes for these deficiencies may be the fact that a 
real system experiences a conformation dependent decrease in the 
entanglement density (or a dilation in the tube diameter) as well as 
convective constraint-release, causing a decrease in the time scale 
for reptation with increasing flow strength that is not included in 
the present form of the DEMG model.  This calls for further efforts 
to improve reptation based constitutive models to obtain a better 
match of its predictions to experiments on entangled polymeric systems. 

\acknowledgments

We thank Johan Remmelgas for helpful discussions, critical reading of 
the manuscript and his help in the flow-symmetry calculation.  We 
acknowledge James P. Oberhauser for the help with the DEMG model numerical 
simulation code.


\narrowtext 

\begin{table}
\caption{The characteristic parameters for the three polystyrene samples.}
\begin{tabular}{cccccccccc} 
Sample & $M_w$ & $c$ &  $\eta_0$ & $N_e$ & $\tau_R$ & $n_t$ & 
$K$ & $n$ \\ 
& ($\times 10^6$) & $(\frac{g}{cc})$ & (P) && (s) && [P $(\frac{rad}{s})^n$] \\ \tableline
PS81 & 8.42 & 0.0396 & 7500 & $\sim 13$ & 2.25 & 8420 
& 343.72 & -0.80 \\
PS82 & 8.42 & 0.0262 & 2700 & $\sim 7$ & 3.01 & 8420 
& 190.11 & -0.70 \\
PS2 & 2.89 & 0.0867 & 19000 & $\sim 13$ & 0.56 & 2890 
& 2703.96 & -0.83 \\
\end{tabular}
\label{samples}  
\end{table} 

\begin{table}
\caption{The characteristic parameters for the two-roll mill with radius $R$, 
height $H$, and gap $G$.}
\begin{tabular}{cccccccccc} 
Roller & $R$ (cm) & $G$ (cm) & $\frac{H}{G}$ & 
$(\frac{\dot{\gamma}_N}{\omega})_{\hbox{th}}$ & 
$(\frac{\dot{\gamma}_N}{\omega})_{\hbox{exp}}$ & $(\lambda_N)_{\hbox{th}}$ & 
$(\lambda_N)_{\hbox{exp}}$ \\ \tableline 
G & 1.278 & 0.844 & 3.1 & 2.7 & 2.8 & 1.501 & 0.1508 \\ 
\end{tabular}
\label{two-roll}  
\end{table} 

\begin{table}
\centering
\caption{The values of the parameter $h(\phi)$ for three orientations $\phi$ 
of the two-roll mill and that for the special case of a symmetric flow 
($\epsilon = 0$).}
\begin{tabular}{ccc} 
$\phi$ & General & Symmetric flow \\ \tableline 
$0^\circ$ & $\dot{\gamma}^2 + \epsilon^2$ & $\dot{\gamma}^2$ \\
$90^\circ$ & $\dot{\gamma}^2\lambda^2 + \epsilon^2$  & $\dot{\gamma}^2\lambda^2$ \\ 
$45^\circ$ & $\epsilon^2 + \frac{1}{2}\dot{\gamma}^2(1 + \lambda^2) + 
\epsilon \dot{\gamma}(1 - \lambda)$ &  $\frac{1}{2} \dot{\gamma}^2 (1 + \lambda^2)$ \\ 
\end{tabular}
\label{tab1}
\end{table}

\end{document}